\documentclass[sigplan]{acmart}

\copyrightyear{2020}
\acmYear{2020}
\setcopyright{acmcopyright}
\acmConference[ASPLOS '20]{Proceedings of the Twenty-Fifth International Conference on Architectural Support for Programming Languages and Operating Systems}{March 16--20, 2020}{Lausanne, Switzerland}
\acmBooktitle{Proceedings of the Twenty-Fifth International Conference on Architectural Support for Programming Languages and Operating Systems (ASPLOS '20), March 16--20, 2020, Lausanne, Switzerland}
\acmPrice{15.00}
\acmDOI{10.1145/3373376.3378522}
\acmISBN{978-1-4503-7102-5/20/03}

\usepackage{wrapfig}
\usepackage{subcaption}
\usepackage{amsthm}
\usepackage{multirow}
\usepackage[normalem]{ulem}
\usepackage[utf8]{inputenc} 
\usepackage[T1]{fontenc}    
\usepackage{url}            
\usepackage{breakurl}

\usepackage{booktabs}       
\usepackage{amsfonts}       
\usepackage{nicefrac}       
\usepackage{microtype}      
\usepackage{graphicx}
\usepackage{textcomp}
\usepackage{amsmath}

\usepackage[font={small,sf,bf}]{caption}

\usepackage{mathtools}

\usepackage{verbatim}
\usepackage{xspace}
\newcommand{\shredder}{\textsf{\small Shredder}\xspace}
\newcommand{\deepmist}{\textsf{\small Shredder}\xspace}

\newcommand{\avgmireducpct}{74.70\%\xspace}
\newcommand{\avgaccloss}{1.58\%\xspace}
\newcommand{\avgconf}{$\pm0.35\%$\xspace}
\newcommand{\avgparams}{$0.16\%$\xspace}
\newcommand{\avgmem}{$19.35$\xspace}
\newcommand{\avgspeedupwifi}{$1.79\times$\xspace}
\newcommand{\avgspeeduplte}{$2.17\times$\xspace}
\newcommand{\avgcompoverhead}{$0.03\times$\xspace}
\usepackage{xparse}
\NewDocumentCommand{\rot}{O{45} O{1em} m}{\makebox[#2][l]{\rotatebox{#1}{#3}}}%
\usepackage{formatting/hadianeh}

\usepackage[subtle]{savetrees}

\usepackage{tikz}
\newcommand*\circled[1]{\tikz[baseline=(char.base)]{
            \node[shape=circle,draw,fill=black,text=white,font=\sffamily\bfseries\small, inner sep=1.2pt] (char) {#1};}}

\newcommand{\niparagraph}[1]{\vspace{2pt}\noindent\textbf{#1}}            
\begin{document}
\fancyhead{}
\title[Shredder]{
Shredder: Learning Noise Distributions to Protect Inference Privacy}
\settopmatter{authorsperrow=1}
\author{Fatemehsadat Mireshghallah\quad Mohammadkazem Taram}
\author{Prakash Ramrakhyani$^\ast$ \quad Ali Jalali$^\dagger$\quad Dean Tullsen\quad Hadi Esmaeilzadeh}

\email{{fmireshg,mtaram}@eng.ucsd.edu,  prakash.ramrakhyani@arm.com, ajjalali@amazon.com, {tullsen,hadi}@eng.ucsd.edu}

\affiliation{%
   Alternative Computing Technologies ({\color[HTML]{0B6121}{ACT}}) Lab\\
   \institution{University of California San Diego\quad\quad\quad $^\ast$Arm, Inc.\quad\quad\quad $^\dagger$Amazon.com, Inc.}
}


\renewcommand{\shortauthors}{Mireshghallah, et al.}
\renewcommand{\authors}{Fatemehsadat Mireshghallah, Mohammadkazem Taram, Prakash Ramrakhyani, Ali Jalali, Dean Tullsen, and Hadi Esmaeilzadeh}
\begin{abstract}
A wide variety of deep neural applications increasingly rely on the cloud to perform their compute-heavy inference.
%
%
This common practice requires sending private and privileged data over the network to remote servers, exposing it to the service provider and potentially compromising its privacy.
Even if the provider is trusted, the data can still be vulnerable over communication channels or via side-channel attacks in the cloud.
%
%
%
To that end, this paper aims to reduce the information content of the communicated data with as little as possible compromise on the inference accuracy by making the sent data noisy.
An undisciplined addition of noise can significantly reduce the accuracy of inference, rendering the service unusable. 
To address this challenge, this paper devises \deepmist, an end-to-end framework, that, without altering the topology or the weights of a pre-trained network, learns additive noise distributions that significantly reduce the information content of communicated data while maintaining the inference accuracy.
The key idea is finding the additive noise distributions by casting it as a disjoint offline learning process with a loss function that strikes a balance between accuracy and information degradation.
%
%
The loss function also exposes a knob for a disciplined and controlled asymmetric trade-off between privacy and accuracy.
While keeping the DNN intact, \deepmist divides  inference between the cloud and the edge device, striking a balance between computation and communication.
In the separate phase of inference, the edge device takes samples from the Laplace distributions that were collected during the proposed offline learning phase and populates a noise tensor with these sampled elements.
Then, the edge device merely adds this populated noise tensor to the intermediate results to be sent to the cloud.
As such, \deepmist enables accurate inference on noisy intermediate data without the need to update the model or the cloud, or any training process during inference.
We also formally prove that \shredder maximizes privacy with minimal impact on DNN accuracy while the tradeoff between privacy and accuracy is controlled through a mathematical knob.
Experimentation with six real-world DNNs from text processing and image classification shows that \deepmist reduces the mutual information between the input and the communicated data to the cloud by \avgmireducpct compared to the original execution while only sacrificing \avgaccloss loss in accuracy.
On average, \deepmist also offers a speedup of \avgspeedupwifi over Wi-Fi and \avgspeeduplte  over LTE compared to cloud-only execution when using an off-the-shelf mobile GPU (Tegra X2) on the edge.

\end{abstract}

\ccsdesc[500]{Security and privacy~Privacy protections}
\ccsdesc[500]{Theory of computation~Nonconvex optimization}
\ccsdesc[500]{Computing methodologies~Neural networks}
\ccsdesc[500]{Networks~Cloud computing}
\ccsdesc[500]{Computer systems organization~Embedded systems}

\keywords{Privacy; neural networks; deep learning; edge computing; cloud computing; inference; noise}

\maketitle

\vspace{-1.5ex}
\section{Introduction}
\label{sec:intro}
Online services that utilize the cloud infrastructure are now ubiquitous and dominate the IT industry~\cite{cusumano2010cloud,cachin2009trusting,motahari2009outsourcing}.
The increasing processing demand of learning models~\cite{jordan2015machine,sze2017efficient} has naturally pushed most of the computation to the cloud~\cite{Hardavellas2011TowardDS}.
Coupled with the advances in machine learning, and especially deep learning, this shift has also enabled online services to offer a more personalized and more natural interface to the users~\cite{Hauswald2015SiriusAO}.
These services continuously receive raw, and in many cases, personal data that needs to be stored, parsed, and turned into insights and actions.
In many cases, such as home automation or personal assistants, there is a rather continuous flow of personal data to the service providers for real-time inference.
While this model of cloud computing has enabled unprecedented capabilities due to the sheer power of remote warehouse-scale data processing, it can significantly compromise user privacy.
When data is processed on the service provider cloud, it can be compromised through side-channel hardware attacks (e.g., Spectre~\cite{spctre} or Meltdown~\cite{meltdown}) or deficiency in the software stack~\cite{heyyou}.
But even in the absence of such attacks, the service provider can share the data with business partners~\cite{Facebook} or government agencies~\cite{23andme}.
Although the industry has adopted privacy techniques and federated learning~\cite{federated} for data collection and model training~\cite{apple17white,erlingsson14ccs}, scant attention has been given to the privacy of users who increasingly rely on online services for inference.

\begin{figure}
    \centering
    \includegraphics[width=0.75\linewidth]{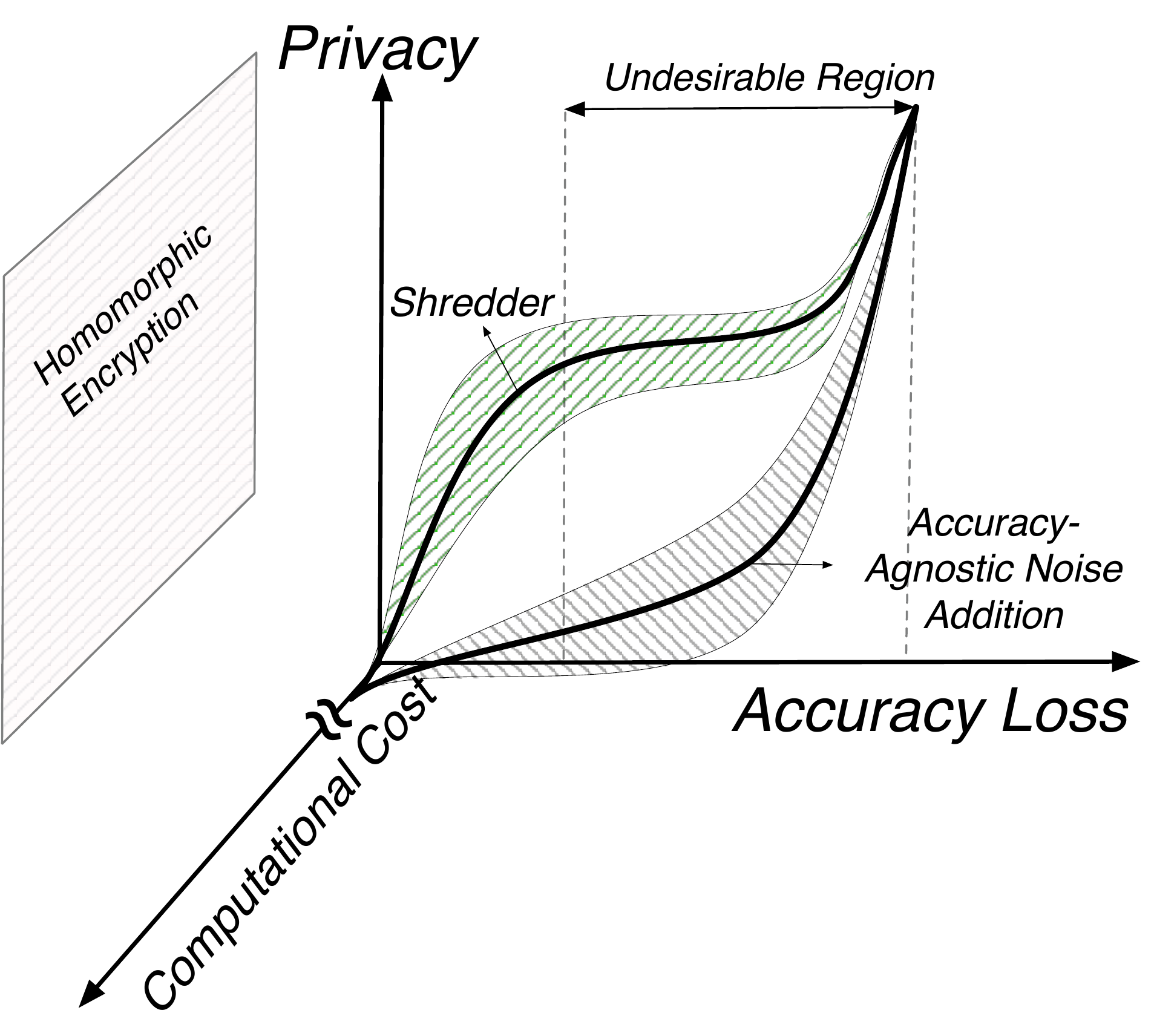}
    \vspace{-1ex}
    \caption{Design space for inference privacy and how \deepmist fits.}
    \label{fig:overview}
\vspace{-1ex}
\end{figure}

As Figure~\ref{fig:overview} illustrates, researchers have attempted to grapple with this problem by employing cryptographic techniques such as multiparty execution~\cite{garbled,smc} and homomorphic encryption~\cite{fullhomomorphic,dowlin16icml,fullhomomorphic2,liu17ccs} in the context of DNNs.
However, these approaches suffer from a prohibitive computation and communication cost, with over three orders of magnitude added latency for CIFAR~\cite{riazi19usenix} for instance, exacerbating the already complex and compute-intensive neural network models.  
Worse still, this burdens additional encryption and decryption layers to the already constrained edge devices despite the computational limit being the main incentive of offloading the inference to the cloud. 

\begin{figure*}
    \centering
    \includegraphics[width=0.9\textwidth]{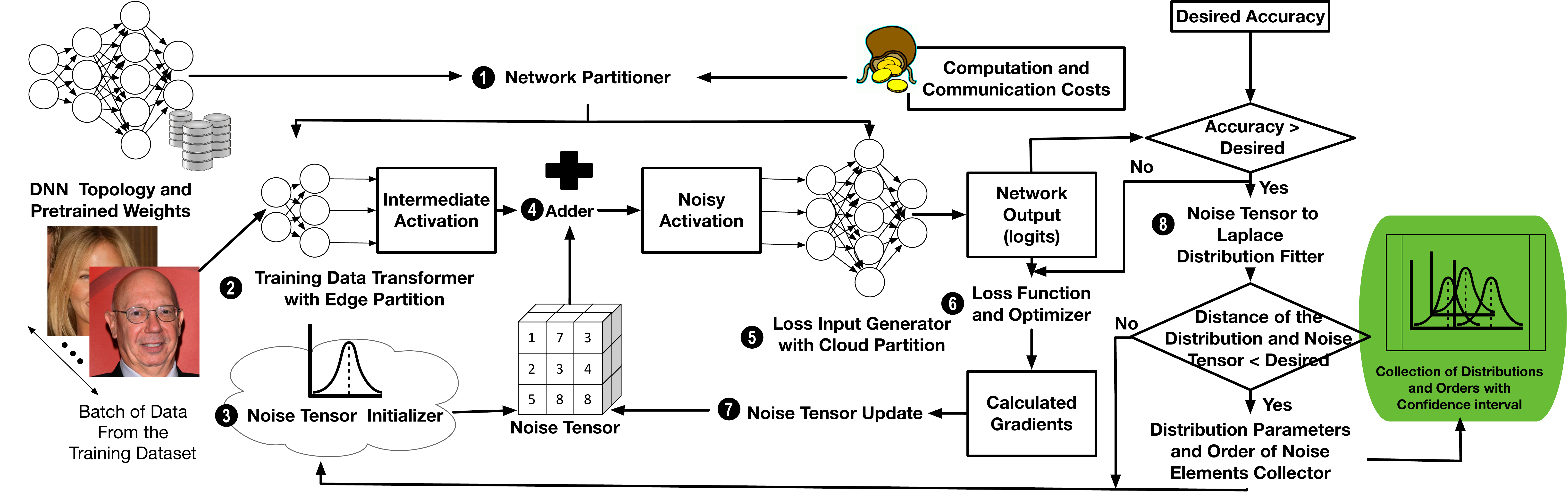}
    \vspace{-1ex}
    \caption{Workflow for offline learning of the noise distribution with \deepmist.}
    \label{fig:overview:train}    
\end{figure*}

This paper, as depicted in Figure~\ref{fig:overview}, takes an orthogonal approach, \deepmist, and aims to reduce the information content of the remotely communicated data through noise injection without imposing significant computational cost.
However, as illustrated, noise injection can lead to a significant loss in accuracy if not administered with care and discipline.
\deepmist resolves this dilemma by finding the noise distributions through a disjoint offline learning process with a loss function that strikes a balance between information loss and accuracy loss.
This loss function also exposes a knob for asymmetrically trading off a modest loss in accuracy for significant improvement in privacy as illustrated in Figure~\ref{fig:overview}.
As such, \deepmist can use the conventional stochastic gradient descent--used for training machine learning algorithms--to learn the noise distributions.

\emph{Our central idea of learning the noise distributions enables \deepmist to mathematically incorporate accuracy as well as the measure of privacy, and if available, the information that is expected to remain private in the loss function.}
The result is a collection of Laplace noise distributions that later during the inference is used by the edge device to scramble the communicated data.
This \emph{offline} process of learning the noise distributions does not require retraining the network weight parameters or changing its topological architecture.
This non-intrusive approach is particularly appealing as most enterprise DNN models are proprietary, and changing a carefully crafted DNN topology and/or its millions of parameters is undesirable.

For inference, \deepmist breaks the DNN between the cloud and the edge device while balancing out computation and communication.
In this separate phase of inference, the edge device takes samples from the Laplace distributions that were collected to populates a noise tensor.
Then, the edge device adds the noise tensor to the intermediate result that is sent over to the cloud.
The number of parameters that \deepmist learns in each epoch matches the number of elements in the intermediate result.
Hence, \deepmist only learns a much smaller collection of parameters--typically <100 KBytes--than the total number of weights--typically >100 MBytes--whose ratio is \avgparams.
As such, the same model can be run on the same cloud on intentionally noisy data without the need for retraining the DNN or the added significant cost of supporting computation on encrypted data.

This problem of offloaded inference is different than the classical differential privacy~\cite{dwork06tcc} setting where the main concern is the amount of indistinguishability of an algorithm.
That is to say, how the output of the algorithm changes if a single user opts out of the input set. 
In inference privacy, however, the issue is the amount of raw information that is sent out.
Nonetheless, since \deepmist employs a Laplace mechanism for generating noise, it is commensurate with differential privacy.
As such, Shanon's Mutual Information (MI)~\cite{shannon1948mathematical} between the user's raw input and the communicated data to the cloud is used as a measure to quantitively discuss privacy. 

We provide a formal formulation and show that \shredder maximizes privacy with minimal impact on DNN accuracy while the tradeoff between privacy and accuracy is controlled through a mathematical knob.
This formal proof is backed by empirical analysis that shows \deepmist reduces the mutual information between the input and the communicated data by \avgmireducpct compared to the original execution with only \avgaccloss accuracy loss over six benchmark networks from text processing and image classification.
It also offers an average speedup of \avgspeedupwifi using Wi-Fi for communication, and \avgspeeduplte using LTE  when an off-the-shelve mobile GPU (Jetson TX2) is used on the edge.

\emph{With these encouraging results the paper contributes \deepmist: 
casting accuracy-aware privacy protection as learning noise distributions through the same algorithm that trains the network, albeit, without retraining the network, while incorporating both privacy and accuracy in its loss.}

\begin{figure*}
    \centering
    \includegraphics[width=0.9\textwidth]{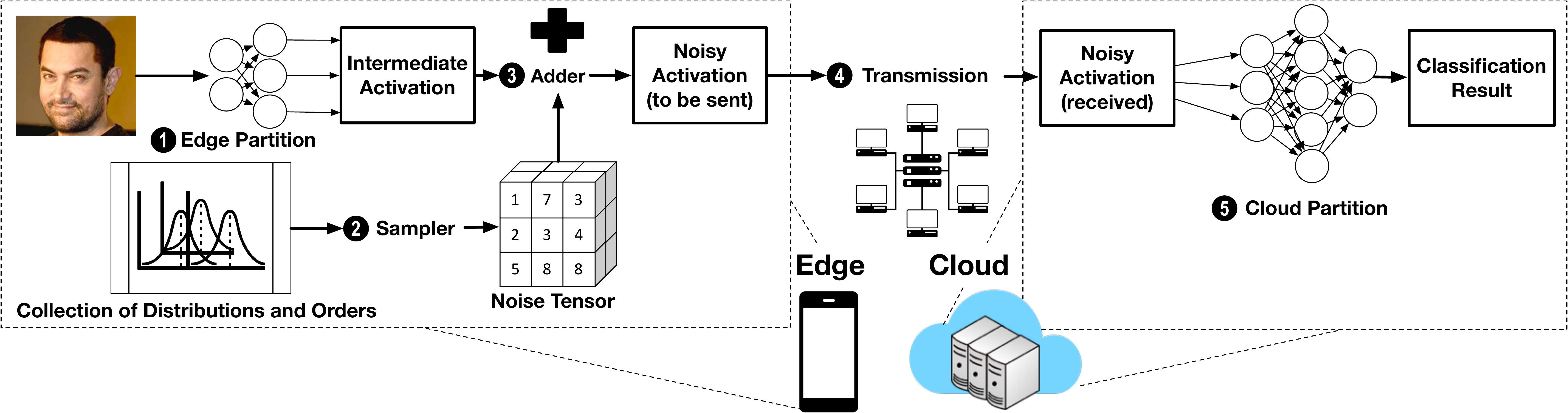}
    
    \caption{Workflow of noisy inference.}
    \label{fig:overview:inf}
    \vspace{-1ex}
\end{figure*}

\vspace{-0.5ex}
\section[short title]{Phase I: Learning the Noise Distributions} 
\label{sec:overv}

\shredder is an end-to-end non-invasive solution that consists of two \emph{disjoint} phases: (1) learning the noise distributions offline and (2) noisy inference.
As Figure~\ref{fig:overview:train} depicts, the first phase takes the DNN topology, its pre-trained weights, a training dataset, the computation and communication costs and acceptable loss in accuracy as inputs. 
The training dataset is the same as the one used to train the DNN.
The output of the phase is a collection of noise tensor distributions coupled with an order for the elements of the tensor for the later phase of inference.
This phase also determines which layer is the optimal choice for cutting the DNN to strike the best balance between computation and communication while considering privacy.
The accuracy and privacy requirements are worked into the mathematics of the learning process.
This section elaborates on the workflow of \deepmist in this first phase, while the next section will discuss the second phase of inference.

\niparagraph{\circled{1} Network partitioner.} 
The first step is to decide where to inject the noise to the DNN.
The partitioner decides the layer at which the neural network should be bisected and offloaded to the cloud. 
This decision is based on the overall computation and communication costs of cutting the network at each layer, and the layer with the lowest cost will be chosen.
The deeper the cutting point, the higher the privacy level, given a fixed level of loss in accuracy. 
This is due to the abstract representation of data in deeper layers of neural networks~\cite{46, 47}.
We also show this observation in our experiments.  
Therefore, to maintain acceptable privacy, the partitioner is set to never cut at the input layer and it should include at least one computational layer.
Nonetheless, there is a trade-off here, between computation costs and privacy, which will be further discussed in Section~\ref{subsec:layer}. 
The partitioning happens only once at the beginning of each learning process and the network is cut to two partitions, \textsf{Edge} and \textsf{Cloud} as depicted in Figure~\ref{fig:overview:train}.
The cut is made such that \textsf{Cloud Partition} starts with a convolution (or fully connected layer), and \textsf{Edge Partition} ends in the layer before that--which could be a pooling layer, or activation functions such as ReLU, depending on the DNN topology.
The rationale is that the pooling layers reduce the data elements and the ReLU layers suppress some of the values, which naturally reduces the information content that is communicated to the cloud.

\niparagraph{\circled{2} Training Data Transformer with Edge Partition.}  
To extract the training (inputs, output) pairs for learning the noise distributions, a batch of the training data is fed to \textsf{Edge Partition}, and the intermediate activation is attained.
These intermediate activations constitute the input part of the pairs for the noise learning process. 
The output part of the pair is obtained in \circled{5}.
Note that in all of the stages of \deepmist, DNN weights are constant and are not altered.

\niparagraph{\circled{3} Noise tensor initializer.} 
Similar to the network partitioner, the noise tensor initializer is executed only once for each learning process. 
It initializes the noise tensor by sampling its elements from an initial Laplace distribution.
The dimension of the noise tensor exactly matches the dimensions of the intermediate activation.  
%

\niparagraph{\circled{4} Adder.} 
The adder adds the noise tensor to the intermediate activation, element-wise.
This stage imitates how the noise will be injected in the future inference runs.

\niparagraph{\circled{5} Loss Input Generator 
with Cloud Partition.}
To be able to go through the noise learning process, a set of outputs is required to pair with the intermediate activation generated in \circled{2}.
To generate the set of outputs, the noisy activations are fed to the \textsf{Cloud Partition}, and the logits (the outputs of the layer before the last classification layer of the neural network, which is usually softmax) are collected.
We use logits because even normal stochastic gradient uses this layer to find the parameters.
The last layer, typically softmax, uses logits to calculate the probability of each class and picks the one which has the highest probability as the classification label.

\niparagraph{\circled{6} Loss function and optimizer.} 
The generated pairs of (input, output) from \circled{2} and \circled{5} are used to calculate the loss, and then fed to an optimizer to calculate the gradients and update the noise tensor using stochastic gradient descent. 
What is important here, is that the gradients are calculated only for the \textsf{Cloud Partition}, and then the adder, but they are not propagated any further.

\niparagraph{\circled{7} Noise tensor update.} 
With the gradients calculated in the previous step,  \textbf{only the noise tensor} is updated, and the DNN weights (even those of \textsf{Cloud Partition}) remain \emph{unchanged}. 
We only calculate the gradients with respect to the weights to be able to update the noise tensor.

\niparagraph{\circled{8} Noise tensor to Laplace distribution fitter.} 
During the noise learning process, after a given number of iterations of training and updating the noise tensor, the accuracy of the model is measured using a held-out set from the initial training dataset. 
If the accuracy is within the desired accuracy given by the user, the noise tensor is fitted to a Laplace distribution.

\niparagraph{\circled{9} Distribution parameters and order of noise elements collector.} 
If the distance between the fitted distribution to the noise tensor and the actual noise tensor is smaller than a pre-determined amount, the parameters of the fitted distribution are saved. 
Also, the descending order of the elements of the noise tensor is saved to be used later during the future inference phase. 
The noise tensor elements, themselves, are discarded and only the order is preserved.
The order is important to preserve the correlation between the elements.
This stage also reports the confidence interval for the accuracy of the model as well.

This workflow is repeated until a certain number of noise distributions are collected.
For our experiments, we collect 20 distributions.
Now that the noise distributions are learned, the DNN can be used by the edge device for inference.
The next section discusses this disjoint process of noisy inference.

\section{Phase II: Noisy Inference} 
Figure~\ref{fig:overview:inf} shows inference with a sampled noise tensor from the learned noise distributions.
This process is very similar to the normal execution of the DNN except that a noise tensor is added to the intermediate activations that are sent to the cloud.
Below, we discuss each step. 

\niparagraph{\circled{1} Edge Partition.} 
User's input data that s/he wants to classify is fed to the \textsf{Edge Partition}.
The output is the intermediate activations.

\niparagraph{\circled{2} Sampler.} 
In parallel with \circled{1}, the Sampler takes in the collection of distribution and orders, i.e. the output of the offline learning phase, and uses that to generate different noise tensors for each inference pass. 
This makes predicting the noise tensor non-trivial for the adversary since for each input data, a different noise is generated \textbf{stochastically}. 
To generate the noise tensor, one distribution is picked randomly from the collection of distributions. 
Then, samples are drawn from this distribution to populate the noise tensor, which has the same dimensions as the intermediate activation.
Then, when the noise tensor is populated, it's elements are rearranged, so as to match the saved order for that distribution. 
For this, the sampled elements are all sorted, and they are replaced according to the saved order of indices in the learning phase.

\niparagraph{\circled{3} Adder.} 
This rearranged noise tensor is simply added to the intermediate activations.
Note that, we do not rearrange the activations and the cloud part of the inference can continue as usual without the cloud knowing about the noise.
The result is the noisy activations.

\niparagraph{\circled{4}  Transmission.} 
At this point, the noisy activation is transmitted over the network from the edge device to the cloud. 
The cloud is completely oblivious to the addition of noise or its nature.

\niparagraph{\circled{5}  Cloud partition.} 
The noisy activation is fed to the \textsf{Cloud Partition}, and the final classification labels are achieved. 
At this point, this result can be sent back to the user or be utilized in the cloud.

\vspace{-1ex}
\section{Noise Distribution Learning Formulation}

This section delves deeper into the details of \shredder, starting from describing the problem formulation and the threat model, and how the trainable noise tensor fits into the context. In addition, this section describes the loss function and training process that \shredder uses for finding the desired noise tensor. We also describe our privacy model and the notions of privacy that we use. 
Our threat model is based on the fact that a large number of real-world DNNs cannot be fully executed on edge devices that range from battery-less devices to more powerful cellphones and tablets.

\vspace{-1.1ex}
\subsection{Threat Model}
Given a pre-trained network $f(x,\theta)$ with $K$ layers and pre-trained parameters $\theta$, we choose a cutting point, $layer_c$, where the computation of all the layers $[0..layer_c]$ are made on the edge. We call this the local network, $L(x,\theta_1)$, where $\theta_1$ is a subset of $\theta$ from the original model. 

\begin{figure}
\centering
\includegraphics[width=0.45\textwidth]{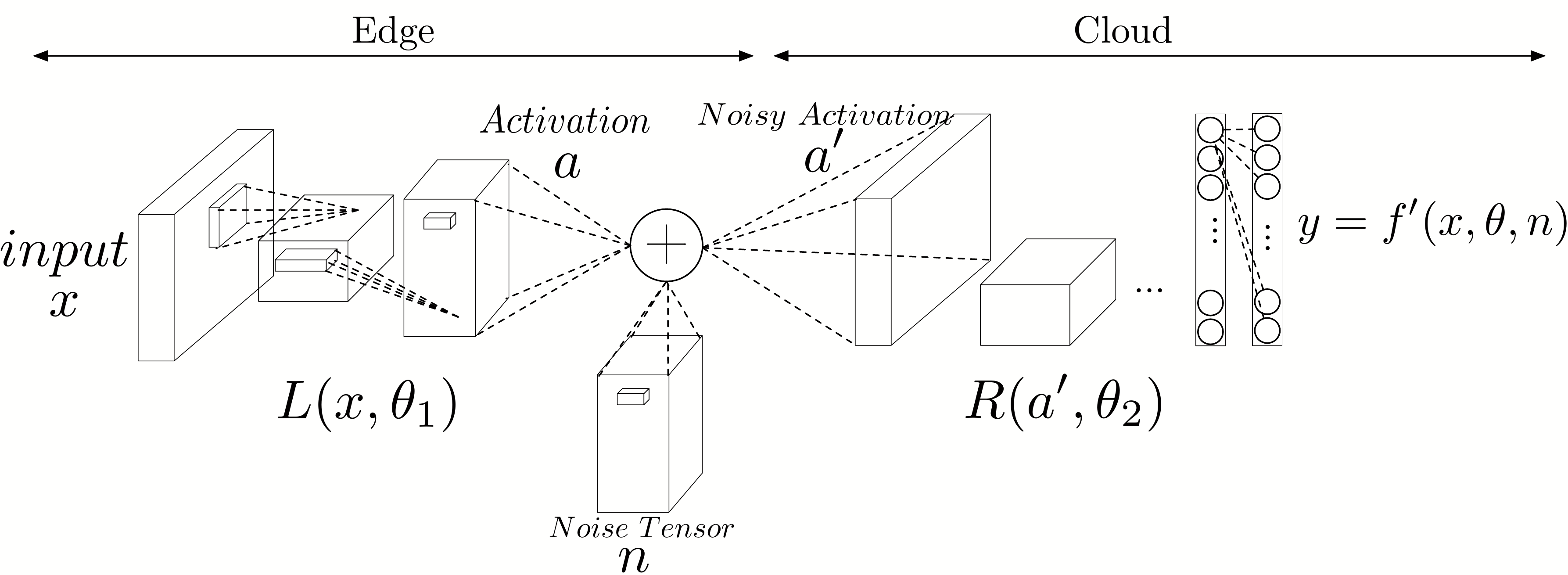}
\caption{ Mathematical modeling of noise injection and learning in \shredder. }
\vspace{-1ex}
\label{fig:proposal}
\end{figure}

The remaining layers, i.e., $[(layer_c+1) .. layer_{K-1}]$, are deployed on the cloud. We call this remote network, $R( x,\theta_2)$. This is shown in Figure~\ref{fig:proposal}. The $f'$ in this image is the noisy network output (noisy logits) which are the outputs of the last layer of the neural network before going through softmax layer.

We assume the cloud is a deep-learning-as-a-service provider to whom users send their primary data($x$) to get the label($y$), where $y = f(x,\theta)$. The threat-model assumes the cloud, or anybody with access to the transmitted data, is untrusted in that it may try to extract information other than $y$ from $x$. We assume the potential adversary(cloud/third-parties) is computationally unbounded. We don’t assume limitations on number of queries to the untrusted cloud and its observations, nor the number of edge-devices that query the service. The local DNN(parameters/model) is known to the cloud/adversary.
The user provides input $x$ to the local network, and an intermediate activation tensor $a = L(x, \theta_1)$ is produced. Then, a noise tensor $n$ is added to the output of the first part,  $a'=a+n$. This $a'$ is then communicated to the cloud where $R(a',\theta_2)$ is computed on noisy data and produces the result $y=f'(x,n,\theta)$ that is sent back to the user.  

The objective is to find the noise tensor $n$ that minimizes our loss function (Section~\ref{sec:train}). To be able to do this through a gradient-based method of optimization, we must find the $\partial y/\partial n$:
\begin{equation}
\begin{split}
    \frac{\partial y}{\partial n} &= \frac{\partial f'(x,\theta, n)}{\partial n} 
                                  = \frac{\partial R((a+n), \theta_2)}{\partial n}  
                                  \\&= \frac{\partial layer_{K-1}}{ \partial Layer_{K-2}} \times ... \times \frac{\partial Layer_{c+1}}{\partial (a+n)} \times \underbrace{\frac{\partial (a +n) }{\partial n}}_{=\frac{\partial n}{\partial n} = 1}
\end{split}
\end{equation}
 Since  $L(x,\theta_1)$ is not a function of $n$, it is not involved in this equation . Gradient of $R$ is also computed through chain rule as shown above. 
 Therefore, the output is differentiable with respect to the noise tensor, which allows for the use of optimization methods like stochastic gradient descent.  
\vspace{-1.2ex}
\subsection{Ex Vivo Notion of Privacy}
%
To measure the privacy, we look at how much information is leaked from input of the network to the data sent across to the cloud. We define information leakage as the mutual information between $x$ and $a$, i.e., $I(x;a)$, where \begin{equation}\label{eq:mi}
    I(x;a) = \int_{x}\int_{a} p_{x,a}log_2{\frac{p_{x,a}}{p_xp_a}}dxda.
\end{equation}

Mutual information has been widely used in the literature for both understanding the behavior of neural networks~\cite{goldfeld18estimating,michael18onbottle,goldfeld18estimating, 47, tishby00information, tishby15bottleneck}, and also to quantify information leakage in anonymity systems in the context of databases~\cite{wang16mi,liao18mi,sankar13mi}. 
We also use the negative of mutual information ($-MI$) as our main and final notion of privacy and call it ex vivo privacy. In our setting, we quantify the information between the user-provided input and the intermediate state that is sent to the cloud. 
\vspace{-1.2ex}
\subsection{In Vivo Notion of Privacy}
%
As the final goal, \shredder\ reduces the mutual information between $x$ and $a'$; however,
calculating the mutual information at every step of the training is too computationally intensive. Therefore, instead, we introduce an in vivo notion of privacy whose whole purpose is to guide our noise training process towards better privacy, i.e, higher $1/MI$. To this end, we use the reverse of signal to noise ratio ($1/SNR$) as a proxy for our ex vivo notion of privacy. Mutual information is shown to be a function of SNR in noisy channels~\cite{guo05additive,guo05tit}. 

\vspace{-1.2ex}
\subsection{Loss Function} \label{sec:train}
\vspace{-1.2ex}
The objective of the optimization is to find the additive noise distribution in such a way that it minimizes $I(x, a')$, the mutual information, and at the same time maintains the accuracy of the primary task. 
In this point, there are two possible scenarios: a) we do not have private labels, which means \shredder's framework is not aware of what it should be protecting against, so it tries to remove any excess information and b) we have private labels, which means \shredder should apply noise which aims at obfuscating that private label, alongside removing other excess information. In the following subsections we will first explain the loss function insight and intuitions with the formulation for the first case, and then build on that and add an extra term to achieve the loss function for the second case.

\subsubsection{No Private Labels Available} \label{subsubsec:primary}

Although the two objectives mentioned previously (minimizing $I(x, a')$, and maintaining accuracy) seem to be conflicting, it is still a viable optimization, as the results suggest. 
The high dimensionality of the activations, their sparsity, and the tolerance of the network to perturbations~\cite{Han2016DeepCC,temam} yields such behavior.

The noise tensor that is added is the same size as the activation it is being added to. The number of elements in this tensor would be the number of trainable parameters in our method. 
\shredder initializes the noise tensor to a Laplace distribution with location parameter $\mu=0$ (location is equivalent to mean of Gaussian distribution). 

We evaluate the privacy of our technique during inference through ex vivo ($-MI$) notion of privacy. However, during training, calculating MI for each batch update would be extremely compute-intensive.
For this reason, \shredder uses an in vivo notion of privacy which uses (SNR) as a proxy to MI~\cite{guo05tit}. In other words, \shredder incorporates  SNR in the loss function to guide the optimization towards increasing privacy.
We use the formulation $SNR =  E[a^2]/\sigma^2(n)$, where $E[a^2]$ is the expected value of the square of activation tensor and $\sigma^2(n)$ is the variance of the noise we add. 
Given the in vivo notion of privacy above, our loss function would be: 
\begin{equation}
    - \sum_{c=1}^{M} y_{o,c}log(p_{o,c}) + \alpha\frac{1}{\sigma^2(n)}
\end{equation}
Where the first term is cross-entropy loss for a classification problem consisting M classes ($y_{o,c}$ indicates whether the observation $o$ belongs to class $c$ and $p_{o,c}$ is the probability given by the network for the observation to belong to class $c$), and the second term is the inverse of variance of the noise tensor to help it get bigger and thereby, increase in vivo privacy (decrease SNR). $\alpha$ is a coefficient that controls the impact of in vivo privacy in training. Since the numerator in our SNR formulation is constant, which is because it is the expected value of activations and it is constant for noisy and original activations across the training dataset,  we do not involve it in the calculations. 
The standard deviation of a group of finite numbers with the range $R=max-min$ is maximized if they are equally divided between the minimum, $min$, and the maximum, $max$. 
This is in line with our observations that show as we push the magnitude of the noise to be bigger, the in vivo privacy would also get bigger.  
Since we initialize the noise tensor with $\mu=0$, some elements are negative and some are positive. The positive ones get bigger, and the negative ones get smaller, therefore, the standard deviation of the noise tensor becomes bigger after each update.  That's why we employ a formulation opposite to L2 regularization~\cite{DBLP:journals/corr/Laarhoven17b}, in order to make the magnitude of noise elements greater. So our loss becomes: 
\begin{equation} \label{eq:loss}
    - \sum_{c=1}^{M} y_{o,c}log(p_{o,c}) - \alpha\sum_{i=1}^{N}{|n_i|}
\end{equation}

This applies updates opposite to L2 regularization term (weight decay and $\alpha$ is similar to the decay factor), instead of making the noise smaller, it makes its magnitude bigger.
The $\alpha$ exposes a knob here, balancing the accuracy/privacy trade-off. In general, as the networks and the number of training parameters get bigger, it is better to make $\alpha$ smaller to prevent the optimizer from making huge updates and overshooting the accuracy. 
 
%

%

%

\subsubsection{Private Labels Available} \label{subsubsec:private}

An example of this case is gender classification and identity classification based on images of faces, which we will discuss more in Section~\ref{sec:ossia}. 
The user may want to classify whether a face is male or female, but s/he does not want the cloud to be able to identify who the person in the image is.  In this case, the primary task is gender classification, and the private task is face identification. 
So, the aim is to minimize the mutual information and maintaining accuracy, similar to Section~\ref{subsubsec:primary}, but with an extra term, to minimize the accuracy of the private task, i.e. face identification. So, we reformulate our loss function to be:
 
 \vspace{-3ex}
 \begin{equation} \label{eq:losspv}
    - \sum_{c=1}^{M} y_{o,c}log(p_{o,c}) + \gamma\sum_{k=1}^{T} z_{o,k}log(p_{o,k}) - \alpha\sum_{i=1}^{N}{|n_i|}
\end{equation}

Where the first term is  cross-entropy loss for the primary classification problem consisting  of M classes ($y_{o,c}$ indicates whether the observation $o$ belongs to class $c$ and $p_{o,c}$ is the probability given by the network for the observation to belong to class $c$), 
the second term is cross-entropy loss for the private classification problem consisting of T classes ($y_{o,k}$ indicates whether the observation $o$ belongs to class $k$ and $p_{o,k}$ is the probability given by the network for the observation to belong to class $k$)
and the last term is the same term from Equation~\ref{eq:loss}, to increase in vivo privacy (decrease SNR).
$\alpha$ is a coefficient that controls the impact of in vivo privacy in training, similar to the previous subsection. $\gamma$ acts similar to $\alpha$, exposing a knob to control the effect of private label accuracy on the overall loss function. 
\subsection{Fitting Noise Tensor to Laplace distribution and Collecting Distribution and Samples} \label{subsec:fit}
During training, whenever the accuracy of the model over the hold-out set of the training dataset exceeds the level desired by the user, the noise tensor is tested for being collected. We use Scipy's \textit{stats} package to fit each learned noise tensor to a Laplace distribution. It's worth mentioning that this stage is executed offline. Then, the probability density function is calculated for the fitted distribution. Using the density function, the distribution collector calculates the Sum of Squared Errors (SSE) between the fitted distribution's histogram and the learned noise tensor's histogram. If the SSE is less than a threshold, the parameters of this distribution are collected. This threshold can be considered as a tunable parameter and differs from model to model and the different desired levels of accuracy and margin of error. 
The element orders of the noise tensor are also saved. By the element orders, the sorted indices of the elements of the flattened noise tensor are meant. For instance, if a tensor looks like $[[3.2, 4.8],[7.3, 1.5]]$, it's flattened version would be $[3.2, 4.8, 7.3, 1.5]$, and the sorted which is what the collector saves would be $[2, 1, 0, 3]$.

\subsection{Loss Function and Noise Learning Analysis} 

\begin{figure}[!t]
      \centering
  \begin{subfigure}{0.35\textwidth}
      \centering
      \includegraphics[width=\textwidth]{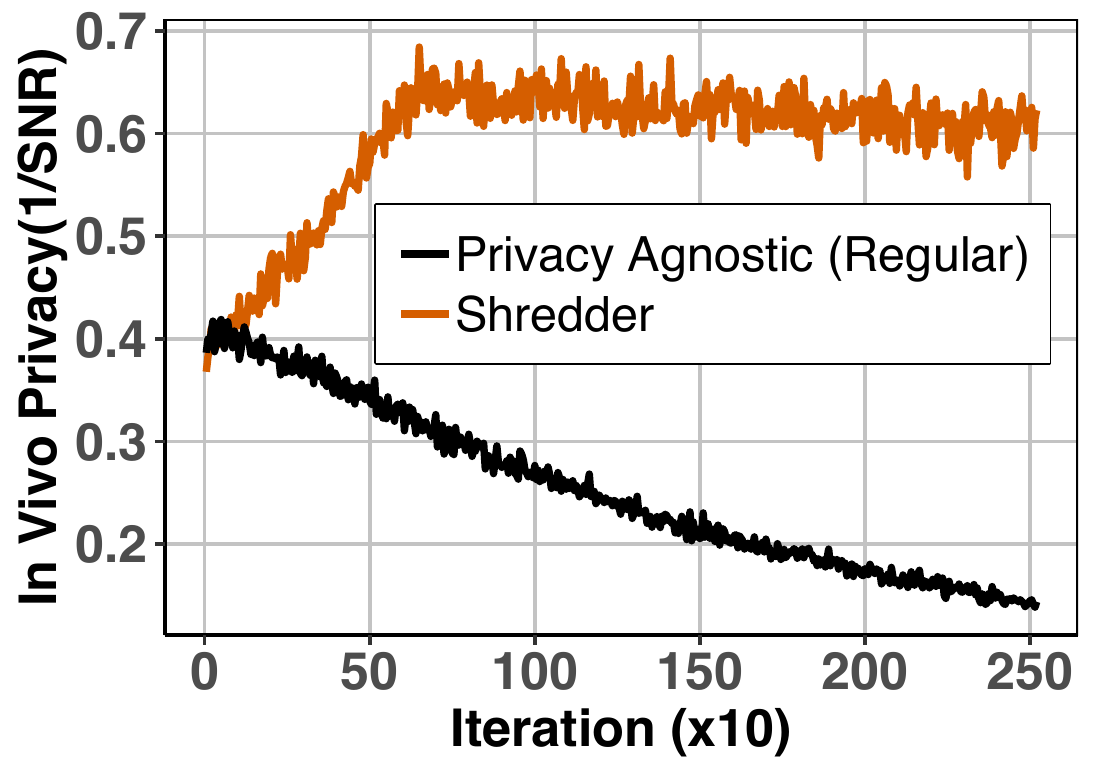} 
  
    \caption{In vivo privacy }
    \label{fig:train:snr} 
  \end{subfigure}
  \begin{subfigure}{0.35\textwidth}
      \centering
      \includegraphics[width=\textwidth]{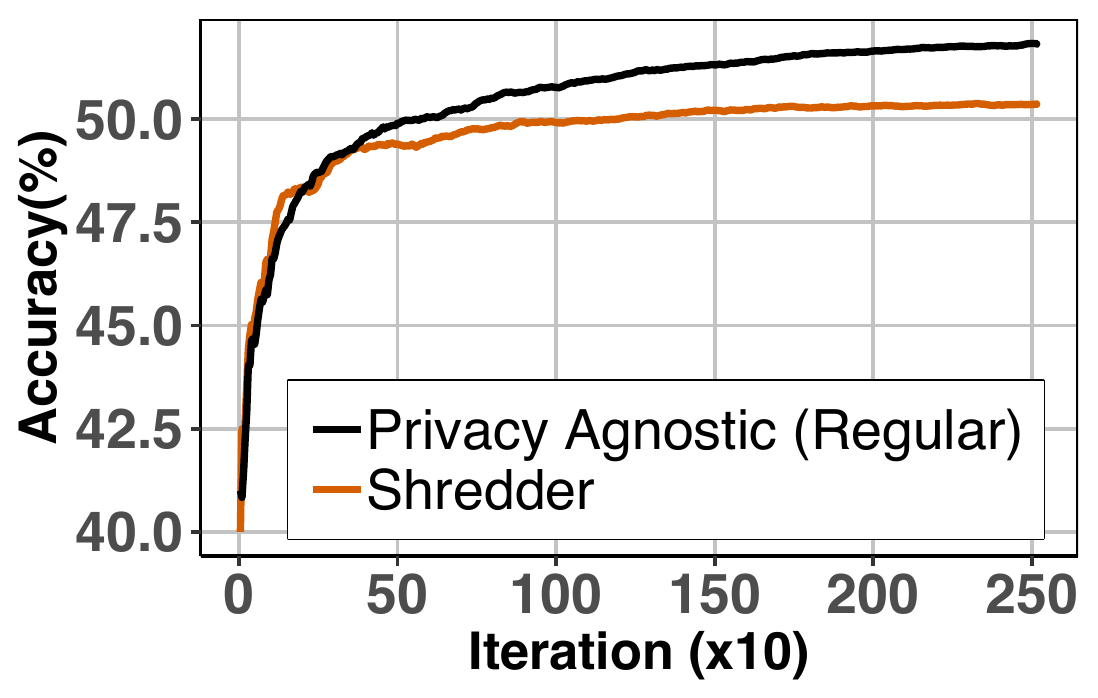} 
    \caption{Accuracy}
    \label{fig:train:acc} 
  \end{subfigure}
  \caption{(a) In vivo notion of privacy and (b) accuracy per iteration of training on AlexNet, when cutting at the last convolution layer. The black lines show regular training with cross entropy loss function. The orange lines show \shredder's learning, with loss function shown in Equation~\ref{eq:loss}.}
  \label{fig:train}

\end{figure}

As Equation~\ref{eq:loss} shows, our loss function has an extra term, in comparison to the regular cross-entropy loss function. This extra term is intended to help decrease the Signal to Noise ratio (SNR). 
Figure~\ref{fig:train} shows part of the training process on AlexNet, cut from its last convolution layer. The black lines show how a regular noise training process would work, with cross-entropy loss and Adam Optimizer~\cite{Kingma2015AdamAM}. 
As Figure~\ref{fig:train:snr} shows in black, the in vivo notion of privacy ($1/SNR$) decreases for regular training  (privacy agnostic, the one without an extra term for decreasing SNR) as the training moves forward. For \shredder\ however, the privacy increases and then stabilizes.

This is achieved through tuning of the $\alpha$ in Equation~\ref{eq:loss}.  $\alpha$ is decayed by 0.1 at every 500 iterations to stabilize privacy and facilitate the learning process. If it is not decayed, the privacy will keep increasing and the accuracy would increase more slowly, or even start decreasing. The accuracy, however, increases at a higher pace for regular training, compared to \shredder\, in Figure~\ref{fig:train:acc}. It is noteworthy that this experiment was carried out on the training set of ImageNet, and when the training is finished, there is negligible degradation in accuracy for \shredder on the test set, in comparison to the regularly trained model.

\vspace{-1ex}
\subsection{Noise Tensor Sampling During Inference}
\vspace{-1.2ex}

This phase is executed during inference, and it samples noise from the collected distributions in the previous phase, as discussed in Section~\ref{sec:overv}. At this point, one of the distributions from the distribution collection (of 20 distributions) is selected randomly. Then, using samples from the chosen distribution, a flattened tensor (a vector) with the size of the noise is populated. The elements of this vector are then re-ordered to match the saved order of indices for that distribution and then reshaped to match the shape of the intermediate activations and get sent to the adder.

%

%

%



\section{Formal Proof and Guarantees}
\label{sec:proof}
This section provides a formal formulation of the \shredder noise learning process and proves that it maximizes privacy with minimal effects on DNN accuracy. 

{\bf Neural Network}: Consider a deterministic function $f(\mathbf{x};\theta^*)\in\mathbb{R}^m$, where, $\mathbf{x}\in\mathbb{R}^p$ is the flatten input to the function and $\theta^*\in\mathbb{R}^r$ is the parameter that fully determines the function. Further, assume that $f(\cdot)$ is a composition of $k$ functions, i.e. $f=f_1\circ f_2 \circ\cdots\circ f_k$, where, $f_i(\cdot;\theta^{*(i)})$ is fully determined by $\theta^{*(i)}$ such that $\theta^*=\left(\theta^{*(1)},\theta^{*(2)},\cdots,\theta^{*(k)}\right)$. In this setting, $f(\cdot)$ represents the trained DNN and $f_i$'s represent network layers. We further define a $\kappa$-split $f = f_L^\kappa \circ f_R^\kappa$ where $f_L^\kappa(\cdot;\theta_L^{*\kappa}) = f_1 \circ \cdots \circ f_{*\kappa}$ and $f_R^\kappa(\cdot;\theta_R^{*\kappa}) = f_{\kappa+1} \circ \cdots \circ f_k$ represent local and remote parts of the network, respectively. We assume DNN is trained and the parameter $\theta^*$ remains the same throughout the process.

{\bf Privacy}: We define the privacy $P_\kappa$ of a $\kappa$-split as the negative of the mutual information between the output of the $i^{th}$ layer and the input, i.e., $P_\kappa=-\mathcal{I}(\mathbf{x};f_L^\kappa(\mathbf{x}))$, where, $\mathcal{I}(\cdot;\cdot)$ represents mutual information. The lower mutual information implies higher privacy. Since $f_L^\kappa$ is a deterministic function, we have 
\begin{equation}
\label{eq:eq1}
	P_\kappa = -\mathcal{H}(f_L^\kappa(\mathbf{x}))+\mathcal{H}\left(f_L^\kappa(\mathbf{x})\big|\mathbf{x}\right) = -\mathcal{H}(f_L^\kappa(\mathbf{x}))
\end{equation}
where $\mathcal{H}(\cdot)$ denotes the entropy function. Moreover, applying Data Processing Inequality, we have $P_1\leq P_2\leq\cdots\leq P_k$, that intuitively means that as we pass the input through more layers, the privacy improves.

{\bf Noise Injection}: Denote a perturbation vector by $\mathbf{w}_\kappa\in\mathbb{R}^{p_\kappa}$ which is independent of $\mathbf{x}$. We intend to perturb the input to $f_R^\kappa$ by changing $\mathbf{y}\!=\!f_L^\kappa(\mathbf{x},\theta_L^\kappa)$ to $\hat{\mathbf{y}}\!=\!f_L^\kappa(\mathbf{x},\theta_L^\kappa)\!+\!\mathbf{w}_\kappa$, i.e., the output of function $f(\cdot)$ changes from $f(\mathbf{x};\theta^*)\!=\!f_R^\kappa(\mathbf{y};\theta_R^\kappa)$ to $\hat{f}(\mathbf{x};\theta)\!=\!f_R^\kappa(\mathbf{y}+\mathbf{w}_\kappa;\theta_R^\kappa)$. It is worth reemphasizing that the parameter $\theta^*$ remains unchanged. This results in a change to the privacy measure as $\hat{P}_\kappa=-\mathcal{I}(\mathbf{x};f_L^\kappa(\mathbf{x})\!+\!\mathbf{w}_\kappa)$. We can provide the following lower bound on the perturbed privacy:
\begin{equation}
\label{eq:eq2}
\begin{split}
\hat{P}_\kappa\;\;\;&\geq -\mathcal{I}\left(f_L^\kappa(\mathbf{x});f_L^\kappa(\mathbf{x})\!+\!\mathbf{w}_\kappa\right)\\
&=-\mathcal{H}(f_L^\kappa(\mathbf{x})) + \mathcal{H}\left(f_L^\kappa(\mathbf{x})\big|f_L^\kappa(\mathbf{x})\!+\!\mathbf{w}_\kappa\right)\\
&=P_\kappa + \mathcal{H}\left(f_L^\kappa(\mathbf{x})\big|f_L^\kappa(\mathbf{x})\!+\!\mathbf{w}_\kappa\right)
\end{split}
\end{equation}
where the last equality is derived from \eqref{eq:eq1}. This equation implies that by noise injection process, we can improve the privacy at least by $\mathcal{H}\left(f_L^\kappa(\mathbf{x})\big|f_L^\kappa(\mathbf{x})\!+\!\mathbf{w}_\kappa\right)$.

{\bf Optimization Problem}: We would like to inject a noise to maximize $\hat{P}_\kappa$ such that the accuracy does not degrade, i.e.,
\begin{equation}
	\mathbf{w}^*_\kappa \;=\; \arg\max_{\mathbf{w}_\kappa}\;\;\;\; \hat{P}_\kappa\;\;\;\;\;\; \mathbf{s.t.}\;\;\;\;\;\mathcal{L}(\hat{f})\leq\mathcal{L}(f)+\epsilon
\end{equation}
where $\mathcal{L}$ is the neural network's loss function. Given \eqref{eq:eq2}, we use $\mathcal{H}\left(f_L^\kappa(\mathbf{x})\big|f_L^\kappa(\mathbf{x})\!+\!\mathbf{w}\right)$ as surrogate objective and reformulate the problem in terms of a Lagrange multiplier $\lambda$ as
\begin{equation}
\label{eq:eq3}
	\mathbf{w}^*_\kappa = \arg\min_{\mathbf{w}_\kappa}\;-\mathcal{H}\left(f_L^\kappa(\mathbf{x})\big|f_L^\kappa(\mathbf{x})\!+\!\mathbf{w}_\kappa\right)+\lambda\mathcal{L}\big(\mathbf{x}, f_R^\kappa(f_L^\kappa(\mathbf{x})+\mathbf{w}_\kappa)\big)
\end{equation}
Denoting $\mathbf{y}\!=\!f_L^\kappa(\mathbf{x})$ and applying the Bayes rule to the conditional entropy, we get $\mathcal{H}\left(\mathbf{y}\big|\mathbf{y}\!+\!\mathbf{w}_\kappa\right)=\mathcal{H}\left(\mathbf{y}\!+\!\mathbf{w}_\kappa\big|\mathbf{y}\right)+\mathcal{H}(\mathbf{y})-\mathcal{H}(\mathbf{y}\!+\!\mathbf{w}_\kappa)=\mathcal{H}(\mathbf{w}_\kappa)+\mathcal{H}(\mathbf{y})-\mathcal{H}(\mathbf{y}\!+\!\mathbf{w}_\kappa)$, where the last equality follows from the fact that $\mathbf{w}_\kappa$ is independent of $\mathbf{y}$. Since $\mathcal{H}(\mathbf{y})$ is constant with respect to $\mathbf{w}_\kappa$, rewrite \eqref{eq:eq3} as
\begin{equation}
\label{eq:eq4}
	\mathbf{w}^*_\kappa = \arg\min_{\mathbf{w}_\kappa}\;\mathcal{H}(\mathbf{y}\!+\!\mathbf{w}_\kappa)-\mathcal{H}(\mathbf{w}_\kappa)+\lambda\mathcal{L}\big(\mathbf{x}, f_R^\kappa(\mathbf{y}+\mathbf{w}_\kappa)\big)
\end{equation}
Optimization problem \eqref{eq:eq4} has three terms. The term $\mathcal{H}(\mathbf{y}\!+\!\mathbf{w}_\kappa)$ controls the amount of information that leaks to the remote part of the DNN and we want to minimize this information. The term $\mathcal{H}(\mathbf{w}_\kappa)$ controls the amount of uncertainty that we are injecting in the form of iid noise. This term is typically proportional to the variance of the noise and we want this term to be maximized (and hence the negative sign). The last term controls the amount of degradation in the accuracy of the DNN and we want to minimize that.

The loss function $\mathcal{L}(\cdot)$ for a $q$-class classification problem with $\mathbf{z}=f(\mathbf{x};\theta^*)\in\mathbb{R}^q$ can be $\mathcal{L}(\mathbf{x},\mathbf{z})=-\sum_{j=1}^q\mathbf{1}_{\mathbf{x}\in \mathcal{C}_j}\log(\mathbf{z}_j)$ where $\mathbf{1}_{\cdot}$ is the indicator function, $\mathcal{C}_j$ represents the $j^{th}$ class and $\mathbf{z}_j$ is the $j^{th}$ entry of $\mathbf{z}$ representing the probability that $\mathbf{x}\in\mathcal{C}_j$. Suppose $\mathbf{1}_\mathbf{x}$ is a one-hot-encoded $q$-vector with $j^{th}$ element being 1 if $\mathbf{x}\in\mathcal{C}_j$ and the rest are zero. We then can write the classification loss in vector form as $\mathcal{L}(\mathbf{x},\mathbf{z})=-\mathbf{1}_{\mathbf{x}}^T\log(\mathbf{z})$. For the remainder of this paper, we target a $q$-class classification problem.

Consider $n$ iid observations of $\mathbf{x}_1,\cdots,\mathbf{x}_n$ where each entry of $\mathbf{y}_i\!=\!f_L^\kappa(\mathbf{x}_i)$ is independently distributed as Laplace distribution $L(\mu_y,b_y)$ and entries of $\mathbf{w}_\kappa$ are iid drawn from $L(0,b_w)$. For Laplace random variables we have $\mathcal{H}(\mathbf{y}+\!\mathbf{w}_\kappa)\propto\log(b_y+b_w)$ and $\mathcal{H}(\mathbf{w}_\kappa)\propto\log(b_w)$. Hence, we can rewrite \eqref{eq:eq4} as an optimization problem on $b_w$ as follows 
\begin{equation}
\label{eq:eq7}
	b_w^* = \arg\min_{b_w}\;\;\log(b_y+b_w)-\log(b_w)+\lambda_n\;\sum_{i=1}^n\mathbf{1}_\mathbf{x}^T\log\big(f_R^\kappa(\mathbf{y}_i+\mathbf{w}_\kappa)\big)
\end{equation}
In order to solve this optimization problem, we start from an initial $\mathbf{w}_\kappa$ and compute the variance of that as $b_w$. We then take a gradient step on $\mathbf{w}_\kappa$ and then update $b_w$ until we converge. This process gives us an optimal $\mathbf{w}_\kappa^*$ and an optimal $b^*_w$.

{\bf Inference}: We first draw $p_\kappa$ (same dimension as the dimension of $\mathbf{w}^*_\kappa$) samples from the Laplace distribution we learned in \eqref{eq:eq7}, i.e., $L(0,b_w^*)$. Then, we sort the samples and create a noise vector $\mathbf{w}_\kappa$ such that for all indices $i$ and $j$, if the $i^{th}$ element of $\mathbf{w}_\kappa^*$ is larger than its $j^{th}$ element, then the same holds for $\mathbf{w}_\kappa$. This process ensures that while we generate random samples, the order of elements are always preserved in $\mathbf{w}_\kappa$.

{\bf Guarantee}: Since in the process of optimizing \eqref{eq:eq7}, we computed the variance from the samples each time, the values of $\mathbf{w}_\kappa^*$ tend to be close to the ordered statistics of the Laplace distribution $L(0,b_w)$. This means that as long as we preserve the order of the values in the noise vector, we statistically stay at the optimal point. This justifies our inference method based on the optimal solution that we get during training.

\section{Evaluation}
\subsection{Methodology} \label{sec:method}

\begin{table}
    \centering
  \caption{Platforms used for obtaining the results in Section~\ref{sec:exp}. We used a TitanXp GPU as our cloud server, and a Tegra X2 GPU (Jetson TX2) as our edge device.}
  \label{tab:platforms}
    \vspace{-3ex}
  \includegraphics[width=0.3\textwidth]{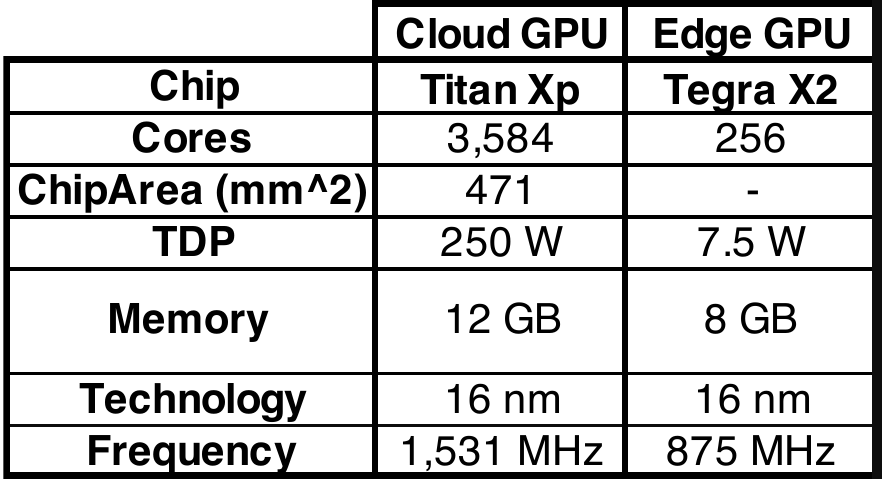}
\end{table}

\begin{table}
    \centering
  \caption{Benchmark networks and datasets used for obtaining the results in Section~\ref{sec:exp}. They are all real-world application networks, used for image classification. The last network/dataset is widely used for text classification in natural language processing.}
  \label{tab:benchs}
  \includegraphics[width=1.0\linewidth]{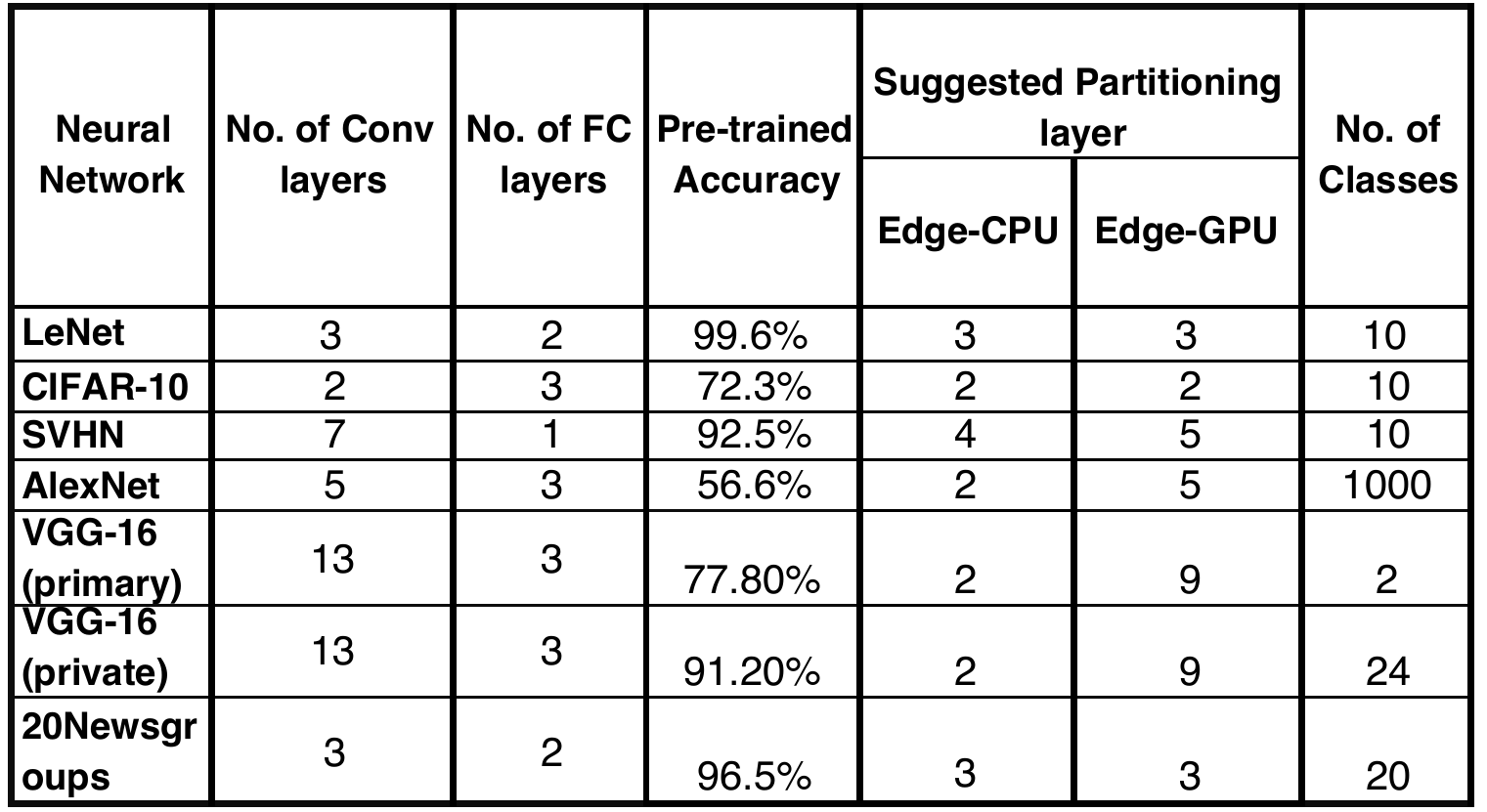}
\end{table}


\paragraph{Benchmarked networks with no private labels.} 
We used 6 real-world application networks as our benchmarks, which can be seen in Table~\ref{tab:benchs}. 
LeNet, CIFAR, SVHN, AlexNet, and VGG-16 are all image classification applications, and  20NewsGroups~\cite{20newsgroups} is a widely used news topic classification application for natural language processing. 
For LeNet, the primary task is digit classification, for CIFAR it is object classification, for SVHN, number street view house number classification, for Alexnet it is image classification into 1000 classes of ImageNet dataset.
VGG-16, the task is gender classification, over a subset of the VGG-Face dataset which has images of celebrities~\cite{vgg-face}.
For the 20Newsgroups, the task is classifying the input text which is extracted from news into 20 topics. 
Besides the aforementioned expected classifications labels, any other information about the image or text is considered private.
For example, for images, what is in the background of the image, if it is day or night, or the identity of the celebrity in the image is considered private. 

\paragraph{Benchmarked networks  with  private labels.} 
For the sake of comparison with the related work~\cite{osia18kde}, we adopt their private labels and primary labels for fairness.
%
%
For the VGG-16 network, we used VGG-Face dataset~\cite{vgg-face}, as seen in Table~\ref{tab:benchs}.   
The private labels for this dataset are the identity of the celebrities in it.
The primary labels are gender classification.

\paragraph{Datasets for experimentation}
Each dataset in Table~\ref{tab:benchs} comes with a training dataset and a validation dataset.
We use the training dataset for learning noise.
During the process of learning, we use a $10\%$ hold-out subset to assess the accuracy.
However, we do not use any of the training datasets for the reported accuracy results.
Instead, we use the validation dataset that is not seen during the learning phase for assessing the final accuracies.

\paragraph{Communication setup.} We have used both private Wi-Fi and AT\&T LTE for communication between edge and cloud.

\paragraph{Network partitioning points.} \shredder can cut the network from any given layer and apply the noise distribution. However, As explained in Section~\ref{sec:overv}, a cutting point that minimizes the overall communication and computation costs is chosen by \shredder. 
Our methodology for modeling the computation and communication costs is commensurate with \cite{neurosergeon}).
The partitioning points used in the experiments can be seen in Table~\ref{tab:benchs}. 
These are the indices of convolution layers, and by cutting at the 3rd layer, we mean the noise is applied right before entering the 4th convolution, and from the 4th convolution to the end of the network, the execution would be on the cloud.

\paragraph{Edge device specifications.}  We used Nvidia Jetson TX2's off-the-shelf GPU  board as our edge device with CUDA V10.0.166 alongside a quad-core ARM A57  that runs  Ubuntu 18.04.2 LTS (GNU/Linux 4.9.140-tegra aarch64).  The specifications of this GPU is in Figure~\ref{tab:platforms}. We used the latest PyTorch version 1.2 built from the source. 

\paragraph{Cloud specifications.}  We used  Nvidia Titan Xp off-the-shelf GPU with CUDA version 10.0.130 with 12GB of RAM  alongside a 12 core Intel Corei9-7920X 2.90GHz CPU that runs Ubuntu 18.04.1 LTS (GNU/Linux 4.15.0-55-generic x86\_64). We used the latest  PyTorch version 1.2  installed with pip3. Table~\ref{tab:platforms} has the specifications of the GPU.

\paragraph{End-to-end cost measurements.} For reporting the cost (computation and communication) of \shredder, we have measured the end-to-end execution time of the DNN on the edge and cloud devices. To be more specific, we have measured the time it takes for the network's first section is executed on the Jetson TX2, then the activations are transferred over Wi-Fi to a Titan Xp server and the rest of network is executed there. 
We have executed 100 times and reported the average. 

\paragraph{Baseline.} We have selected cloud-only execution as our baseline since it is the widely used approach for edge devices~\cite{wide1, wide2}.

\paragraph{Offline noise distribution learning phase setup.} For this, we use the same hardware and software setup as the cloud.  

\paragraph{Privacy measurement setup.} Mutual Information (MI) is calculated using the Information Theoretical Estimators Toolbox's~\cite{szabo14information} Shannon Mutual Information with KL Divergence. 
In the results reported in upcoming sub-sections, MI is calculated over the shuffled test sets on MNIST~\cite{lecunnMNIST} dataset for LeNet~\cite{LeCun1998GradientbasedLA},
CIFAR-10 dataset for CIFAR-10~\cite{Krizhevsky2010ConvolutionalDB},  SVHN dataset for SVHN~\cite{Goodfellow2014MultidigitNR},  ImageNet~\cite{ILSVRC15} dataset for AlexNet~\cite{Krizhevsky2012ImageNetCW}, a subset of VGG-Face~\cite{vgg-face} for the VGG-16 dataset and the 20Newsgroups~\cite{20newsgroups} dataset for 20Newsgroup's neural network. These photos were shuffled through and chosen at random.
Using mutual information as a notion of privacy means that \shredder targets the average case privacy, but does not guarantee the amount of privacy that is offered to each individual user.

\paragraph{$\alpha$ and $\gamma$ parameters.} 
For the parameters (knobs) mentioned in Equations~\ref{eq:loss} and ~\ref{eq:losspv}, We use $-0.01$, $-0.001$,  and $-0.0001$ for $\alpha$  on \{LeNet, CIFAR, 20Newsgroups\}, \{SVHN\}  and \{AlexNet, VGG-16\} respectively. For $\gamma$, we use 0.01 for VGG-16, since it is the only network with defined private labels.

\paragraph{Optimizer setup for offline noise distribution learning.} We used the cross-entropy loss function with Adam Optimizer ~\cite{Kingma2015AdamAM}, with a learning rate of 0.01 on LeNet, CIFAR  and 20Newsgroups, 0.001 on SVHN and 0.0001 on AlexNet and VGG-16.

\paragraph{Comparison with DPFE setup.} 
We compare \shredder with DPFE\cite{osia18kde} over VGG-16 network, for face identification and gender classification on celebrity faces, which is the exact setup used by~\cite{osia18kde}   to evaluate their method. DPFE offers only this benchmark, and we have used the same network partitioning point as well, to provide fairness. We have used celebrity faces from the VGG-Face dataset and partitioned it to validation/training portions to train and verify the classifiers.

\vspace{-1.3ex}

\subsection{Experimental Results}
\label{sec:exp}

This section elaborates on our observations in detail and brings empirical evidence for the efficacy of \shredder. We discuss the memory footprint and latency overheads of \shredder, accuracy-privacy trade-off, a comparison with another method, and finally, a network partitioning point privacy and edge computation trade-off analysis. For all the experiments, except the ones in Section~\ref{sec:ossia}, the loss function form of Equation~\ref{eq:loss} has been used since no assumption on the private labels is given, which is the case in most privacy-related problems. In Section~\ref{sec:ossia}, we assume private labels (Equation~\ref{eq:losspv}) and compare  \deepmist with and without the private labels against  related work, DPFE~\cite{osia18kde}.

Table~\ref{table:epochs} summarizes our experimental results. We have indicated the margin of error for each benchmark here. 
Since the margin is trivial, we have only brought it in the table, and not in the plots, for the sake of simplicity. It is shown that on the networks, \shredder can achieve on average \avgmireducpct loss in information while inducing \avgaccloss loss in accuracy. 
The table also shows that it takes \shredder a short time to train the noise tensor, for instance on AlexNet on ImageNet and 1000 classes, it is 0.1 epoch. It is also evident that the memory overhead of \shredder due to its distribution collector is insignificant.

\begin{table*}[t]
\small
\sffamily
\centering
\caption{Summary of the experimental results of \shredder for the benchmark networks. }
\begin{tabular}{@{}llllllll@{}}
\toprule  
Benchmark  & LeNet     & CIFAR     & SVHN      & Alexnet & VGG-16& 20Newsgroups & Average\\\midrule
Original Mutual Information  &  301.84   &  236.34   &  19.2     & 12661.51&28732.21 & 27.8  & --  \\      
Shredded Mutual Information  &  18.9     &   90.2    &  7.1      &  4439.0  & 7268.7 & 7.8  & --  \\
\textbf{Mutual Information Loss}      &  \textbf{93.74\%}  &  \textbf{61.83\%}  &  \textbf{64.58\%}  &  \textbf{64.94\%}&  \textbf{74.70\%} & \textbf{72.95\%} & \textbf{\avgmireducpct } \\
\textbf{Accuracy Loss}                &  \textbf{1.34\%}   &   \textbf{1.42\%}  &  \textbf{1.12\%}   &  \textbf{1.95\%} &  \textbf{1.68\%} & \textbf{1.99\%}   &\textbf{\avgaccloss}  \\
Margin of Error               &  \textbf{$\pm0.39\%$}   &   \textbf{$\pm0.87\%$}  &  \textbf{$\pm0.22\%$}   &  \textbf{$\pm0.12\%$} & \textbf{$\pm0.11\%$}   & \textbf{$\pm0.41\%$}   & \textbf{\avgconf}  \\
\shredder's Learnable Params over DPFE\cite{osia18kde} &0.19\%&0.65\%& 0.04\%    &  0.02\%&  0.01\%& 0.05\%   & \avgparams  \\         
Number of Epochs of Training        &  6.3      &  1.7      &  1.2      & 0.1  &  6.8  &  7.3 & 1.99  \\
\shredder's Distribution Collector Memory Footprint (KB)  &  2.05     &   8.70    &  5.00     &  315.00  & 918.75 &  2.01& 19.35  \\

\bottomrule
\end{tabular}
\label{table:epochs}

\end{table*}

\begin{figure}[!t]
    \vspace{1ex}
    \centering
        \includegraphics[width=0.45\textwidth]{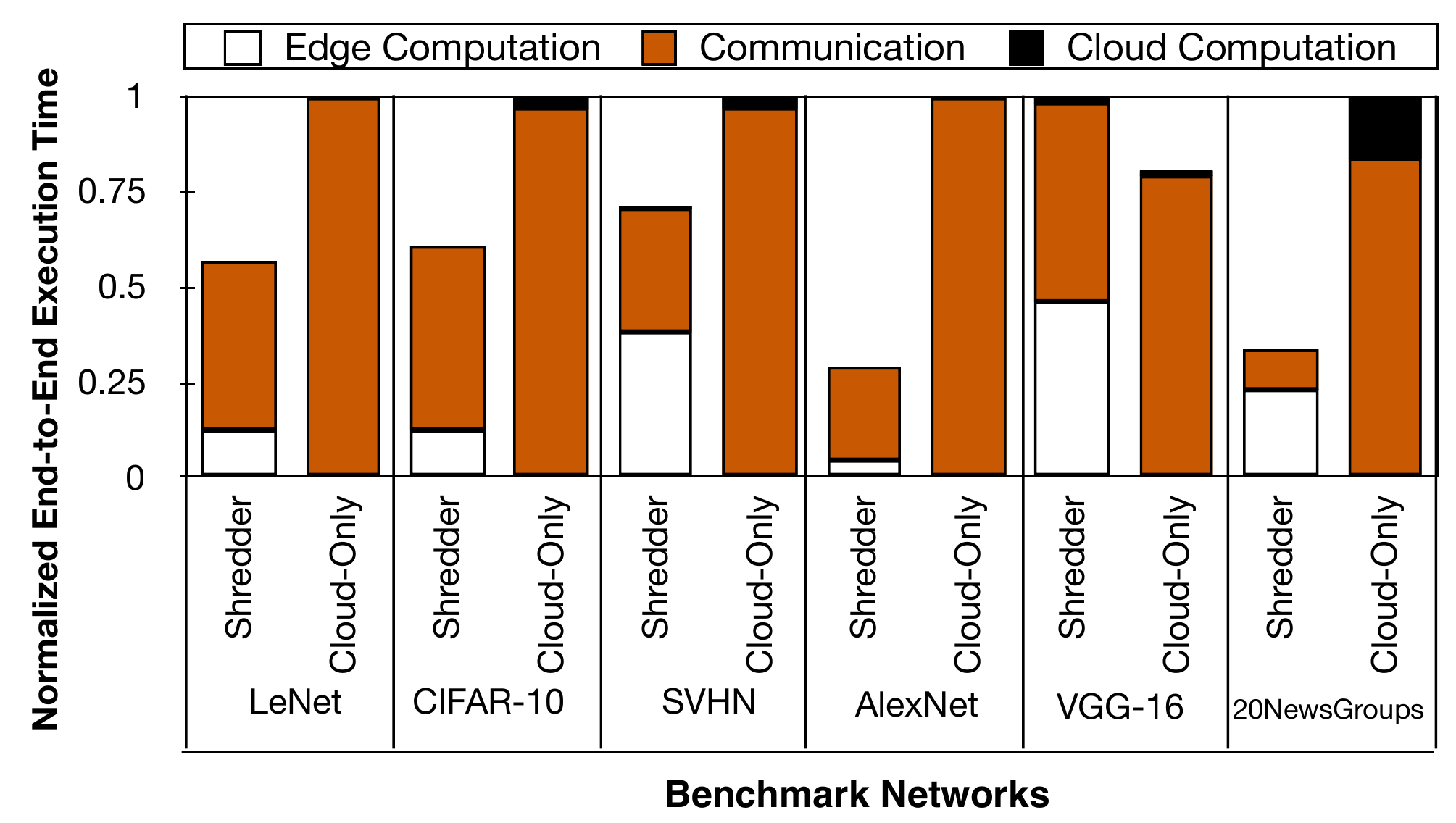} 
        
        \label{fig:overhead} 

    \caption{Comparison of \shredder's end-to-end execution time, compared to a baseline of cloud-only execution which is widely used. In this evaluation the partitioning points are those mentioned in Table~\ref{tab:benchs} and Wi-Fi is used for communication.}
    \label{fig:overheads}
    \vspace{0ex}
\end{figure}

\vspace{-1ex}
\subsubsection{Overheads of \shredder}

As Table~\ref{table:epochs} shows, \shredder alleviates the complexities of training an entire neural network, by decreasing the number trainable parameters of the network to \avgparams. 
The collection of distributions and element orders imposes a trivial memory footprint of saving the distribution parameters (two numbers per distribution, which are the $location$ and $scale$, are similar to mean and standard deviation) and the orders of tensor elements for each distribution, which is a string of numbers. The length of this string is the same as the size of the noise tensor, and each element is smaller than or equal to the length of the string, since the elements are the sorted indices of the noise tensor, as explained in Section~\ref{subsec:fit}. There are 20 distributions collected for each benchmark. The average of all these overheads combined is \avgmem KBs. 

Figure~\ref{fig:overheads} shows the end-to-end execution time of \shredder, normalized to a baseline of cloud-only execution. The Figure shows that \shredder outperforms the baseline for all benchmarks, except VGG-16 and it has an average speedup of \avgspeedupwifi. These results are all commensurate with Neurosergeon and show the same trends~\cite{neurosergeon}. 
This speedup is mainly because of the high communication costs and the large size of the input to the network that should be transmitted to cloud for cloud-only execution, whereas \shredder sends a much smaller intermediate layer. It is important to keep in mind that the computation overhead of the distribution sampler and noise adder is on average \avgcompoverhead the communication time of \shredder, and is therefore trivial in comparison to the communication costs.  

The reason for VGG-16's slow-down is that for this network, over Wi-Fi, the optimal cutting point is the input layer, and since \shredder does not partition network from input layer to provide more privacy, it partitions from the convolution 13th layer which is the second optimal computation and communication point, and at this point, the activation size is still relatively big, so, it does not help compensate the excess computation time imposed to the edge device by the execution of the previous 13. Therefore, the overall execution time is increased by $23\%$, in comparison to the cloud-only approach. 

As mentioned, communication is measured over Wi-Fi, which is faster than LTE and 3G, so \shredder's speedup would be even higher if one of the slower communication methods were employed. For instance, if LTE is used, the overall speedup would be \avgspeeduplte and VGG-16 would have a speedup of $1.03\times$.


\vspace{-1.1ex}
\subsubsection{Accuracy-Privacy Trade-Off}
\begin{figure*}[!t]
    \centering
  \begin{subfigure}{0.25\textwidth}
      \centering
      \includegraphics[width=\textwidth]{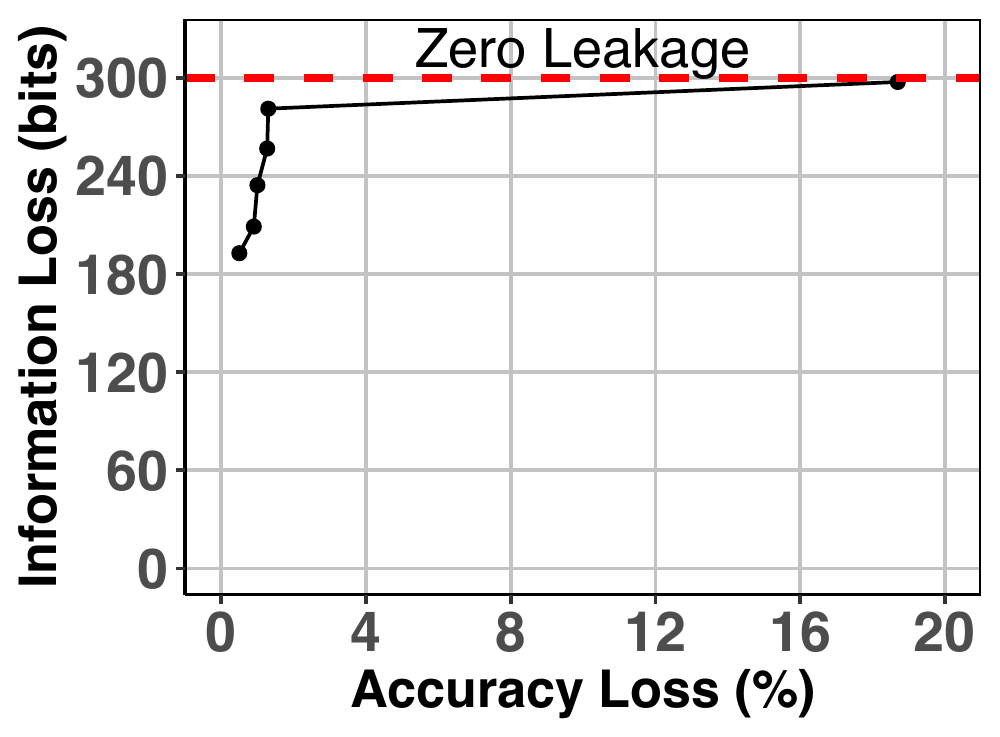} 
    \vspace{-1.4em}
    \caption{LeNet}
    \label{fig:priv:lenet} 
  \end{subfigure}
  \begin{subfigure}{0.25\textwidth}
      \centering
      \includegraphics[width=\textwidth]{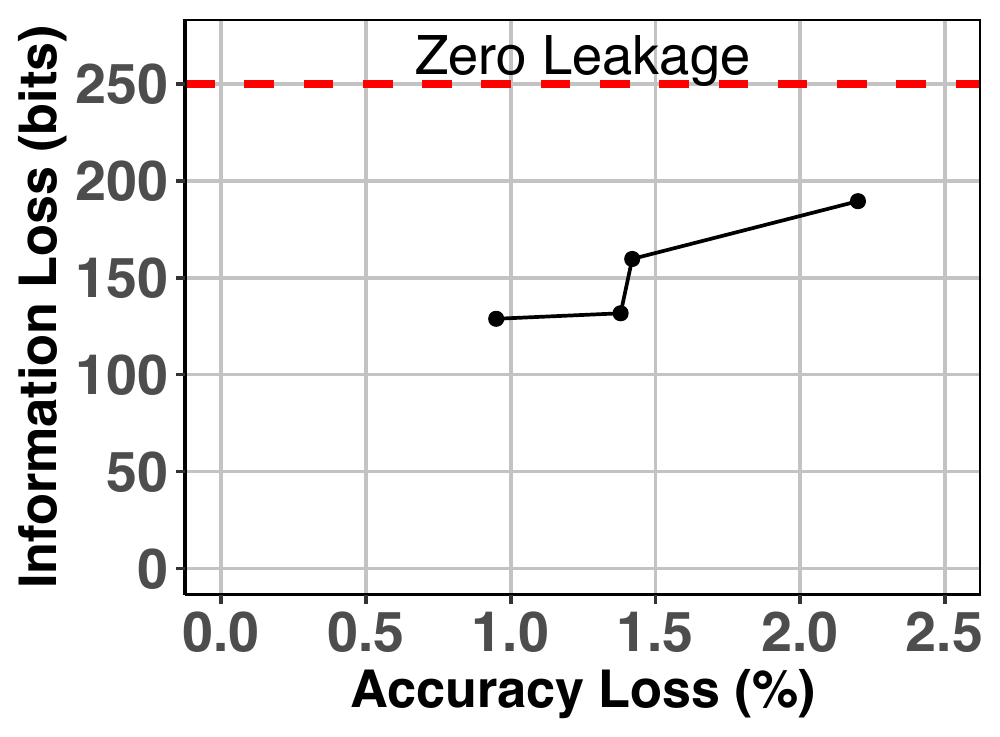} 
    \vspace{-1.4em}
    \caption{CIFAR}
    \label{fig:priv:cifar} 
  \end{subfigure}
  \begin{subfigure}{0.25\textwidth}
      \centering
      \includegraphics[width=\textwidth]{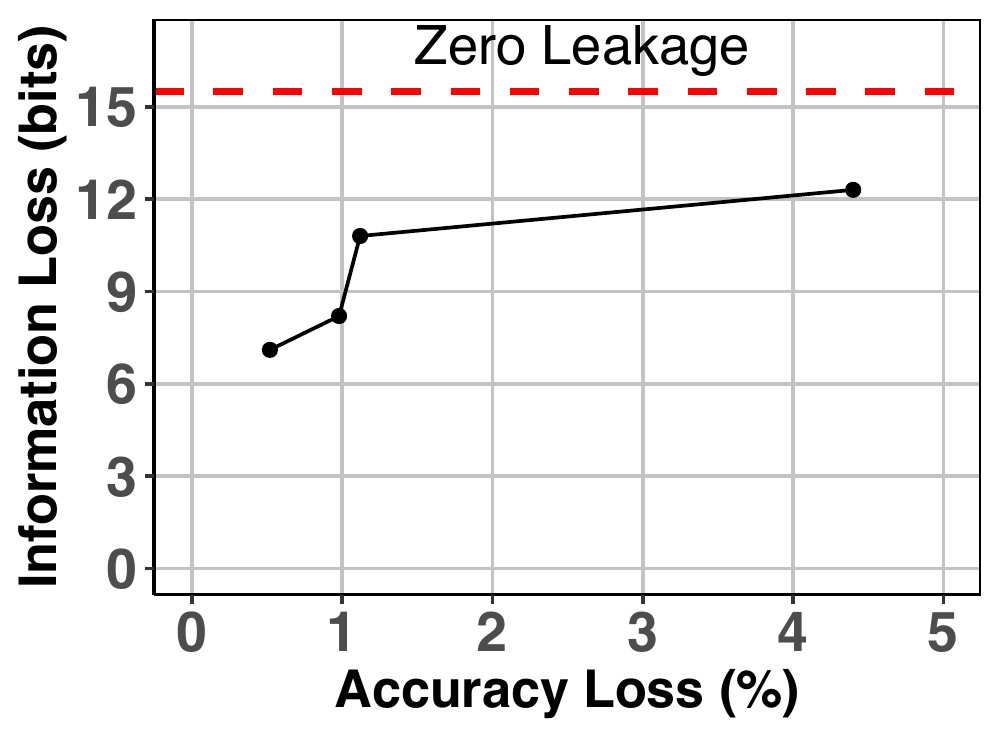} 
    \vspace{-1.4em}
    \caption{SVHN}
    \label{fig:priv:svhn} 
  \end{subfigure}

  \begin{subfigure}{0.25\textwidth}
    \centering
    \includegraphics[width=\textwidth]{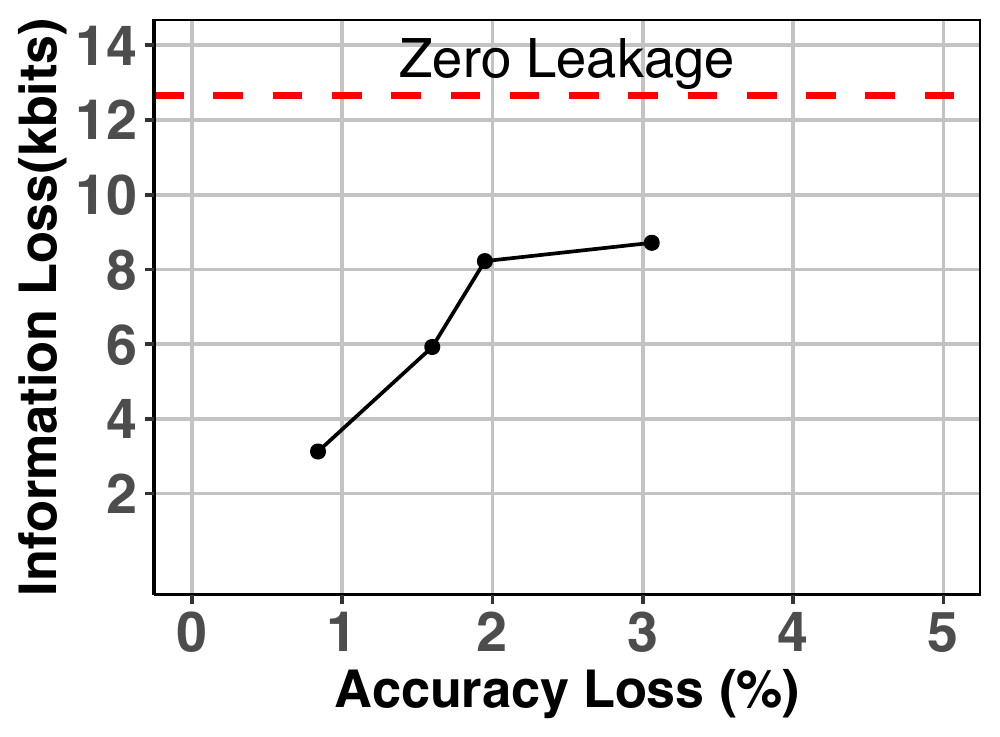} 
    \vspace{-1.4em}
    \caption{AlexNet}
    \label{fig:priv:alexnet} 
\end{subfigure}
 \begin{subfigure}{0.25\textwidth}
    \centering
    \includegraphics[width=\textwidth]{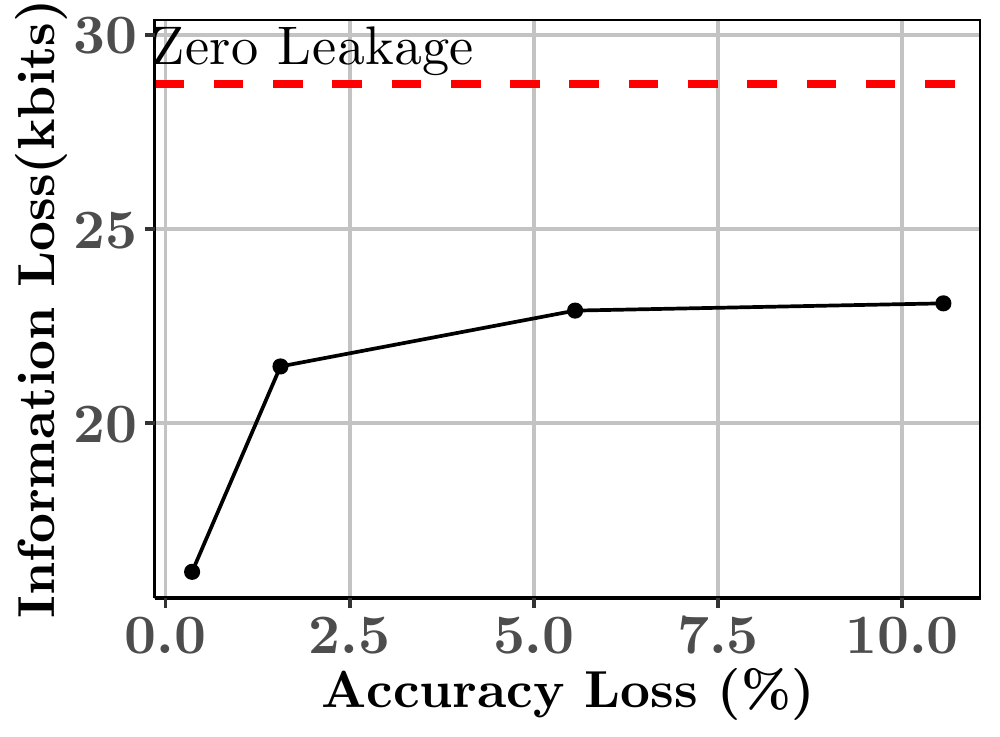} 
    \vspace{-1.4em}
    \caption{VGG-16}
    \label{fig:priv:vgg16} 
\end{subfigure}
  \begin{subfigure}{0.25\textwidth}
    \centering
    \includegraphics[width=\textwidth]{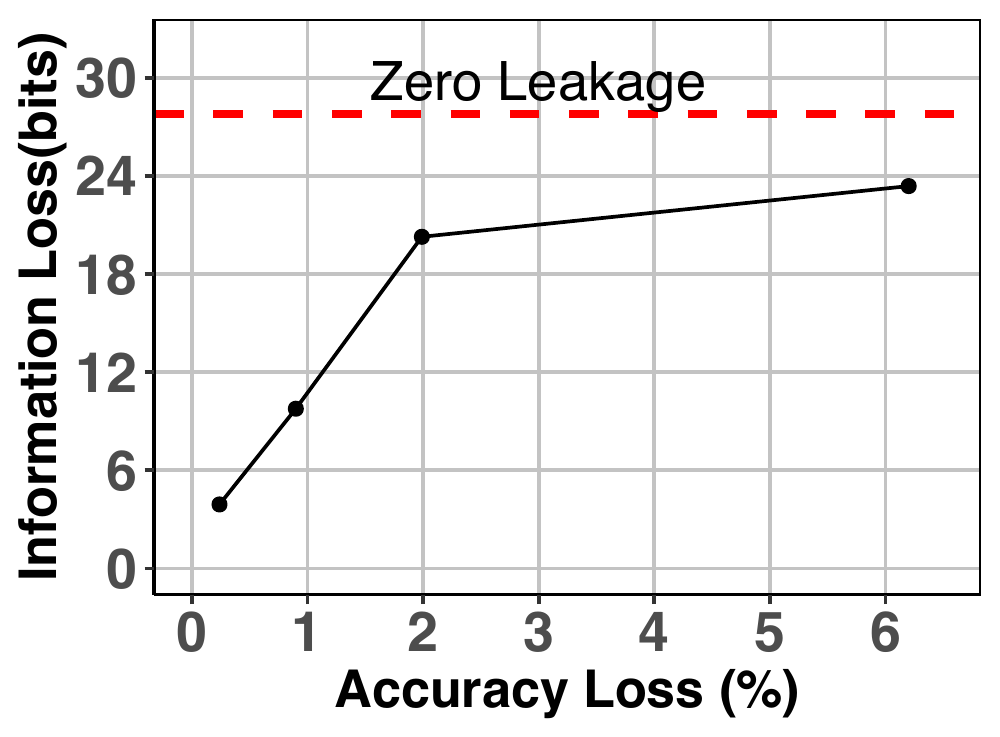} 
    \vspace{-1.4em}
    \caption{20Newsgroups}
    \label{fig:priv:newsgroups} 
\end{subfigure}

  \caption{Accuracy-Privacy trade-off in 6 benchmark networks. The zero leakage line shows the original mutual information between input images and activations at the cutting point.}
  \label{fig:priv}

\end{figure*}

 There is a trade-off between the amount of noise that we incur to the network and its accuracy.  
As shown in Figure~\ref{fig:overview}, \shredder\ attempts to increase privacy while keeping the accuracy intact. Figure~\ref{fig:priv} shows the level of privacy that can be obtained by losing a given amount of accuracy for LeNet, CIFAR-10, SVHN, and AlexNet. 
In this Figure, the number of mutual information bits that are lost from the original activation using our method is shown on the $Y$ axis. The cutting point of the networks is their last convolution layer. This can be perceived as the output of the \textit{features} section of the network, if we divide the network into \textit{features} and \textit{classifier} sections. 
%

The \textit{Zero Leakage} line depicts the amount of information that needs to be lost to leak no information at all. In other words, this line points to the original number of mutual information bits in the activation that is sent to the cloud, without applying noise. 
The black dots show the information loss that \shredder\ provides, given a certain loss in accuracy. These trends are similar to that of Figure~\ref{fig:overview}, since \shredder\ tries to strip the activation from its excess information, thereby preserving privacy and only keeping the information that is used for the classification task. This is the sharp (high slope) rise in information loss, seen in sub-figures of Figure~\ref{fig:priv}.
%
%
Once the excess information is gone, what remains is mostly what is needed for inference. That is why there is a point (the low slope horizontal line in the figures) where adding more noise (losing more information bits) causes a huge loss in accuracy. The extreme to this case can be seen in~\ref{fig:priv:lenet}, where approaching the \textit{Zero Leakage} line causes about $20\%$ loss in accuracy. 

\begin{figure}[!t]
    \vspace{1ex}
    \centering
    \begin{subfigure}{0.49\textwidth}
        \centering
        \includegraphics[width=\textwidth]{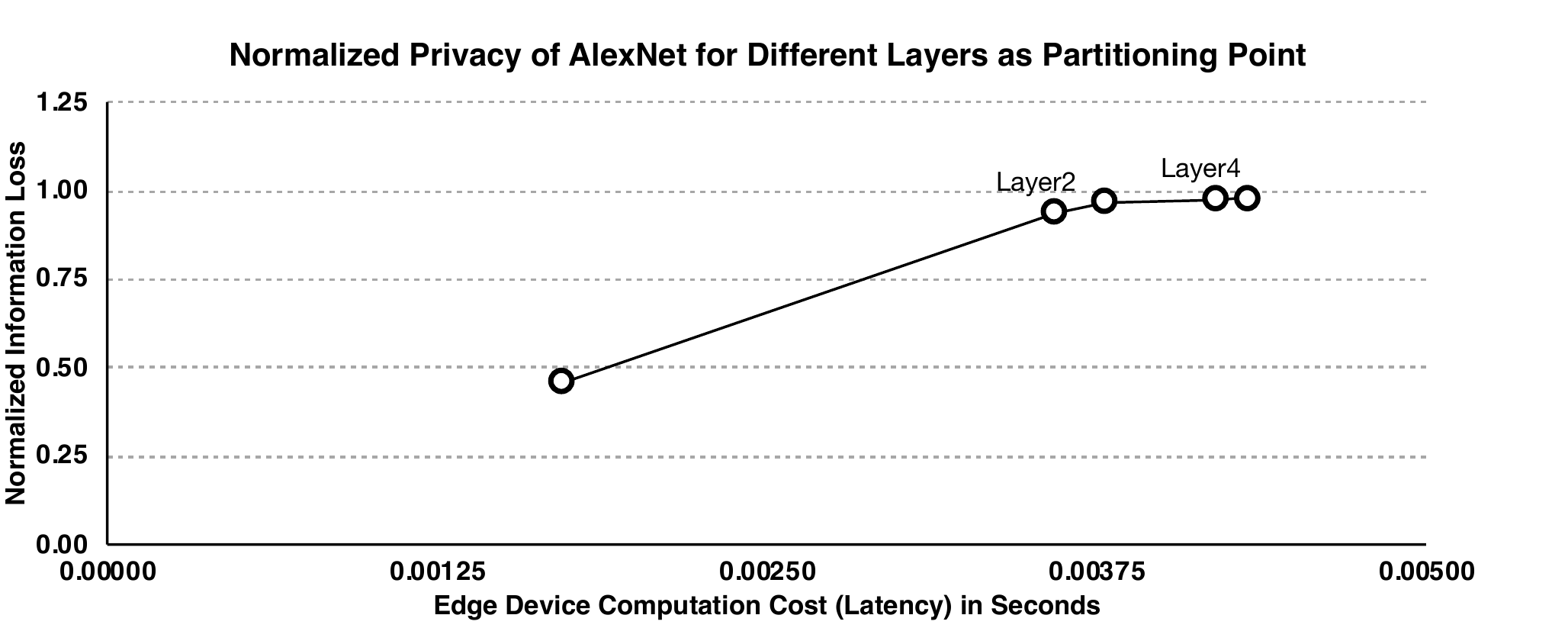} 
        
        \caption{Normalized privacy over last layer }
        \label{fig:alex:priv} 
    \end{subfigure}
    \begin{subfigure}{0.49\textwidth}
        \centering
        \includegraphics[width=\textwidth]{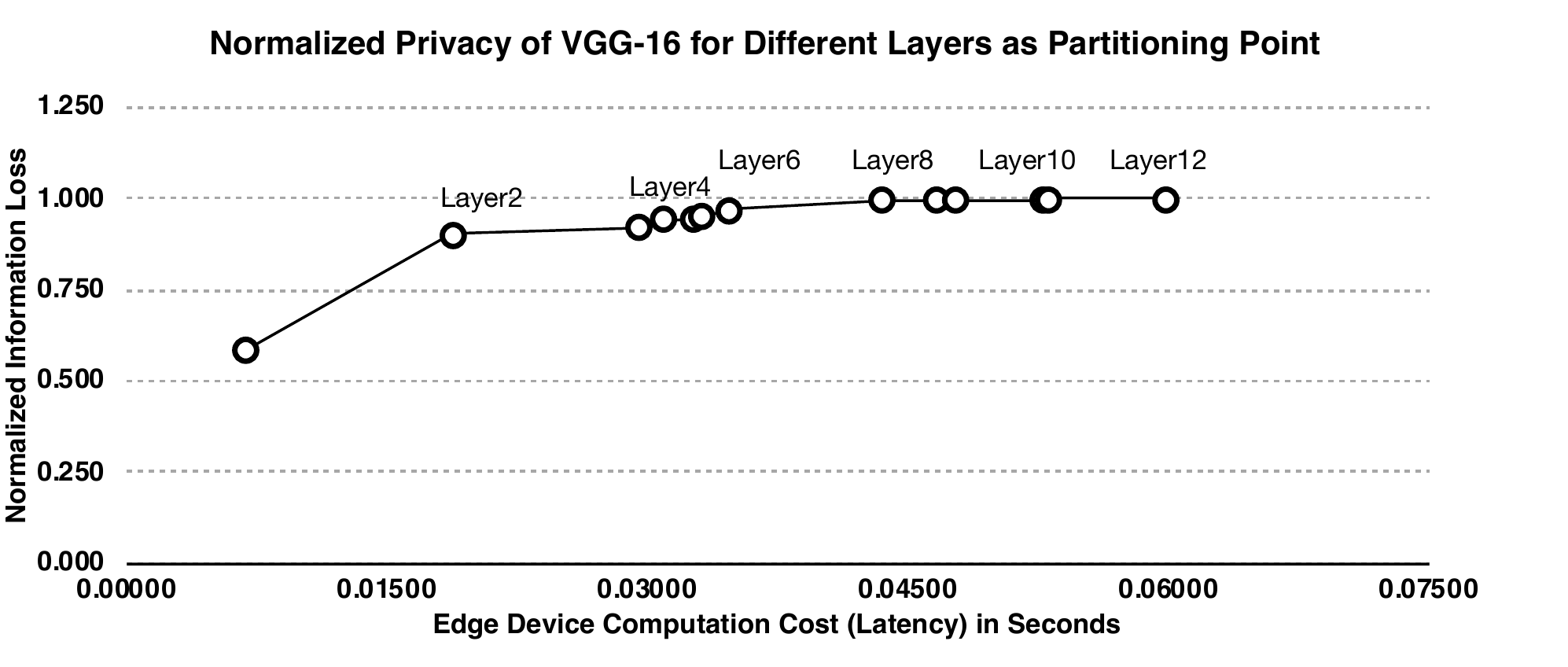} 
        \caption{Normalized  privacy over last layer}
        \label{fig:vgg:priv} 
        
    \label{fig:priv:cut:shallow}    
    \end{subfigure}

    \caption{Normalized  privacy over last layer for different levels of computation on the edge device (different partitioning points) for (a) AlexNet on ImageNet and (b) VGG-16 on VGG-Face dataset. After a point (3rd convolution layer for AlexNet and 8th convolution layer for VGG-16) the improvements in privacy start to diminish.}
    \label{fig:priv:comp}
    \vspace{0ex}
\end{figure}

\vspace{-1.1ex}
\subsubsection{Partitioning Point Trade-offs}
\label{subsec:layer}

Layer selection for network cutting point is mostly an interplay of communication and computation costs of the edge device. As discussed in Section~\ref{sec:overv}, the network partitioner chooses the layer with the lowest end-to-end execution cost. The deeper this layer is, the higher privacy yield for a given accuracy~\cite{osia18kde}, since the network operations like pooling layers, ReLU and convolutions themselves modify the input information and give the \shredder framework a lower mutual information to begin with~\cite{46, 47}. 
That's why the partitioner it set to never choose the input layer as partitioning point (which is what cloud-only execution is similar to) since it compromises privacy. However, if the user wants even higher privacy, they could cut the network deeper and get higher privacy, at the cost of more edge computation. But the relation is not linear, and the returns for privacy start diminishing at some point. Figure~\ref{fig:priv:comp} shows the highest normalized privacy that can be reached with less than $5\%$ loss in accuracy for different edge device computation time (different partitioning layers) over AlexNet and VGG-16. The numbers on the $x$ axis show the computation time and each point on the plot shows a layer. For AlexNet, after the 3rd convolution layer, and for VGG-16, after the 8th one, the increase in privacy is insignificant as we move forward through the convolution layers. The trend is similar for other networks, for the sake of space, we have chosen VGG-16 and Alexnet as representative of other DNNs~\cite{Coates2013DeepLW}. 
 The users can weigh the trade-offs, and \textbf{sacrifice costs for more privacy} and cut deeper layers. However, as we show in the experiments, the privacy plateaus at some point and is not ever-increasing. The suggested partitioning points can be seen in Table~\ref{tab:benchs}


%


%

\begin{figure}[!t] 
    \centering
    \begin{subfigure}{0.38\textwidth}
        \centering
        \includegraphics[width=\textwidth]{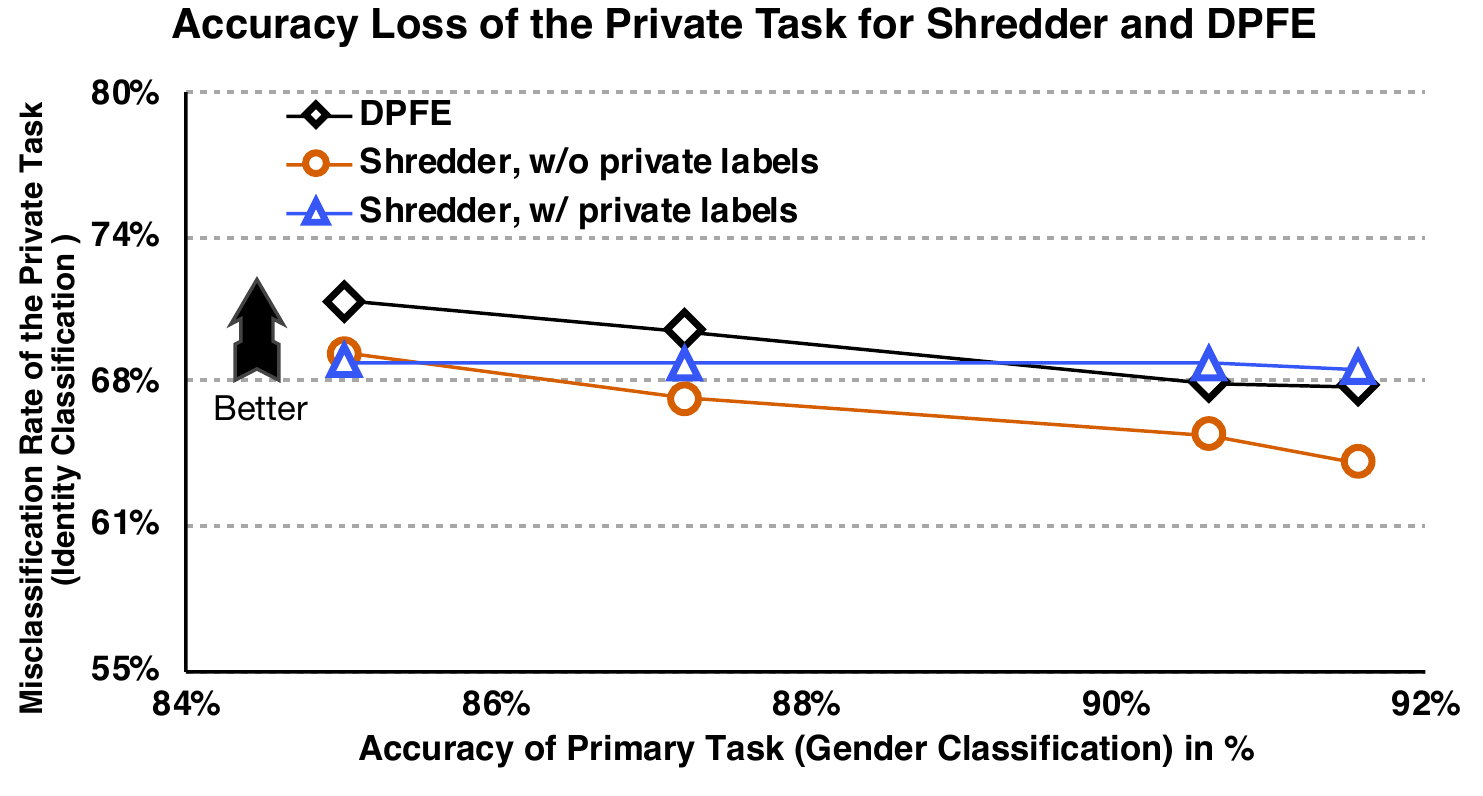} 
        \vspace{-1.4em}
        \caption{Misclasification rate of private labels}
        \label{fig:ossia:acc} 
    \end{subfigure}
    \begin{subfigure}{0.38\textwidth}
        \centering
        \includegraphics[width=\textwidth]{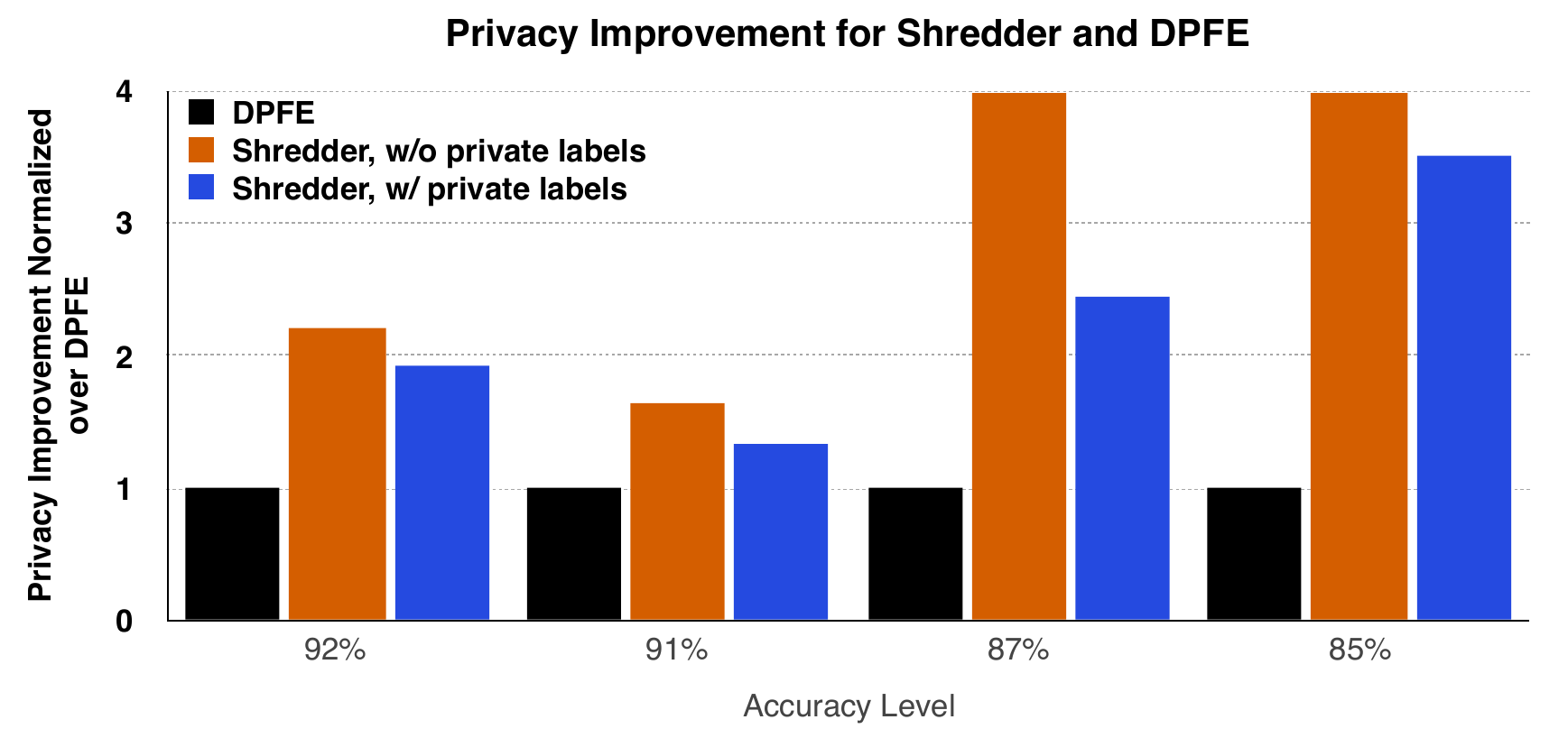} 
        \caption{Normalized privacy improvement over DPFE}
        \label{fig:ossia:priv} 
    \end{subfigure}

    \caption{Misclasification rate of private labels (identity) and privacy improvement comparison for different accuracy levels of the primary task (gender classification) over VGG-16 on VGG-Face dataset, for \shredder with both loss functions and DPFE.}
    \label{fig:ossia}
    
\end{figure}

\subsubsection{Comparison with DPFE}\label{sec:ossia}

Deep Private Feature Extraction (DPFE)~\cite{osia18kde} is a privacy protection mechanism that aims at obfuscating given private labels, by modifying the network topology and re-training all the model parameters. 
DPFE partitions the network in two partitions, first partition to be deployed on the edge and the second on the cloud. It also modifies the network architecture by adding an auto-encoder in the middle and then re-training the entire network with its loss function.  
DPFE's loss function is composed of three terms, first, the cross-entropy loss which aims at maintaining accuracy, second, a term that tries to decrease the distance between intermediate activations with different private labels, and a final term which tries to increase the distance between intermediate activations of inputs with the same private label. After training, for each inference, a randomly generated noise is added to the intermediate results on the fly.

DPFE can only be effective if the user knows what s/he wants to protect against, whereas \deepmist offers a repetitively more general approach that tries to obliterate any information that is irrelevant to the primary task.  Table~\ref{table:epochs} has a row that compares the number of trainable parameters for \deepmist with DPFE and it can be seen that \deepmist's parameters are extremely lower than DPFE's. 

To run experiments, the intermediate outputs of the networks are fed to two classifiers, for gender and identity, each of which display an original (before adding noise) accuracy of $91.56\%$ and $77.8\%$, respectively.  Then DPFE and \deepmist are applied to the neural networks, and the accuracy results over validation sets are seen in Figure~\ref{fig:ossia}.

Figure~\ref{fig:ossia:acc} compares DPFE to \deepmist with its Equation~\ref{eq:loss} (without private labels) and Equation~\ref{eq:losspv} (with private labels) in terms of the misclassification rate of the private task, i.e. the identity classification (identity compromise)  for given levels of accuracy for the main task,  which is the same metric used in~\cite{osia18kde}.  As expected, \deepmist without private labels has the lowest misclassification rate, which is due to not having any knowledge of the private data. At higher levels of primary accuracy, \deepmist with private labels outperforms DPFE. This can be attributed to \deepmist's loss function, especially the last term, which aims at maximizing the amount of noise while keeping the accuracy intact. This could be the reason why \deepmist has a higher misclassification rate in a more constrained setting (higher primary accuracy). Whereas DPFE adds the noise on the fly, which can only help increase the misclassification rate to a certain degree. 
In lower accuracy levels where the state-space is less constrained, DPFE offers higher misclassification, which can be attributed to it's a higher number of trainable parameters which can give it more wiggle room. 

Figure~\ref{fig:ossia:priv}  shows the privacy improvement of \deepmist with both loss functions, over DPFE. It can be inferred from the Figure that \deepmist without the private labels performs better since it has a more general approach which scrambles more overall information. However, \deepmist with the private labels and also DPFE, take an approach which is directed at a specific goal, which impedes them from providing privacy for aspects other than the private task. This is seen more in DPFE than \deepmist with private labels since the latter still tries to maximize noise standard deviation. 
%

%


\vspace{-1ex}
\section{Related Work}
%
The literature abounds with a variety of attempts to provide greater protection to users' private data in a neural processing system~\cite{wei,shokri15ccs,abadi16DeepLW,liu17ccs,dowlin16icml}. 
These efforts span different levels of the system, from training to inference. 
The majority of these studies~\cite{shokri15ccs,abadi16DeepLW}; however, have focused on preserving the privacy of contributing users to statistical databases or training models. ~\cite{fed-priv}, for instance, introduces a privacy-preserving protocol for federated learning.
These techniques tackle the inherent conflict of extracting useful information from a database while protecting private or sensitive data of the individuals from being extracted or leaked~\cite{dwork14book}.
As Table~\ref{tab:related} illustrates, the landscape of research in privacy for neural networks can be categorized into the efforts that focus on training or inference. 
These categories can be further grouped according to whether or not they require retraining the DNN weights or modifying the model itself (i.e., intrusive).
\shredder falls in the category of the techniques that are non-intrusive and target the inference phase.

The other technique in this same category, MiniONN~\cite{liu17ccs}, uses homomorphic encryption that imposes non-trivial computation overheads making it less suitable for inference on edge.
Below, we discuss the most related works, which typically require obtrusive changes to the model, training, or add prohibitively large computation overheads.  

\paragraph{Adding noise for privacy.} 
The idea of noise injection for privacy goes back at least to the very first differential privacy papers~\cite{dwork06euro,dwork06tcc} where they randomize the result of a query to a database by adding noise drawn from a Laplace distribution.
More recently, Wang et al.~\cite{wang18kdd} proposes data nullification and noise injection for private inference.

However, unlike \shredder, they retrain the network.
Osia et al.~\cite{osia18kde,osia2017hybrid} design a private feature extraction architecture that uses principal component analysis (PCA) to reduce the amount of information. 
Leroux et al.~\cite{leroux2018privacy} use an autoencoder to obfuscate the data before sending it to the cloud, but the obfuscation they use is readily reversible, as they state. 
We, on the other hand, cast finding the noise as differentiable noise tensor while considering accuracy in the loss function of the optimization that finds the noise.

\paragraph{Trusted execution environments.} 
Several research propose running machine learning algorithms in  in trusted execution environments such as Intel SGX \cite{mckeen13sgx} and ARM TrustZone \cite{alves04trustzone}  to address the same remote inference privacy~\cite{tramer2018slalom,hunt2018chiron,hanzlik2018mlcapsule,olga17usenix} as well as integrity~\cite{tramer2018slalom}. 
However, the privacy model in that research requires users to send their data to an enclave running on a remote servers.
In contrast to \shredder, this model still allows the remote server to have access to the raw data and as the new breaches in hardware~\cite{spctre, meltdown, ZombieLoad, foreshadow,foreshadowNG} show, the access can lead to comprised privacy.

\paragraph{Differential privacy.} 
As a mathematical framework, differential privacy~\cite{dwork06euro,dwork06tcc,dwork14book}  was initially proposed to quantify privacy of users in the context of privacy-preserving data-mining or statistical databases.
To this end, it measures the degree to which the algorithm behaves similarly if an individual record is in or out of the database/training set. 
This definition gives a robust mathematical guarantee to the question of -- given a private training set (or, database entry) as input, how safe is the trained model (or, aggregate database) to publish~\cite{jdif}. 
Naturally, differential privacy has also been employed in training of deep neural networks~\cite{shokri15ccs, abadi16DeepLW} where the datasets may be crowdsourced and contain sensitive information.  
The research on differential privacy is largely in \textit{centralized models}, where users trust a curator who has access to the whole pool of private data~\cite{dwork14book}. 
In a more practical model, called local differential privacy, the system does not require users to even trust the curator to inspect their data, even for the purpose of preserving privacy~\cite{bittau17sosp,erlingsson14ccs,ding17nips,apple17white}.
In this setting, which the system is just collecting data and not performing inference, the data is still scrambled on the edge devices.
This scrambled data is then remotely aggregated and just provides an average trend across multiple sources.
The existing differential privacy models are in fact solving a fundamentally different problem than \shredder.
They are concerned with data collection while \shredder aims to improve privacy during a real-time cloud-enabled inference.

\textbf{Encryption and cryptographic techniques.} 
Secure multiparty computation (SMC)~\cite{garbled,smc} and homomorphic encryption~\cite{liu17ccs,dowlin16icml,juvekar18usenix} have also been used as attempts to deal with the privacy on offloaded computation on the cloud~\cite{dowlin16icml, chabanne17iacr,riazi19usenix,chabanne17iacr,juvekar18usenix, liu17ccs, mohassel17sp}.
Secure multiparty computation refers to a group of protocols that enable multiple parties to jointly compute a function while each party solely has access to its own part of the input~\cite{mohassel17sp,smc}.
To establish trust and isolation, SMC relies on compute-heavy encryption or obfuscation techniques.
To adopt SMC to the privacy problem, recent works~\cite{mohassel17sp} assume a two-party secure computation in which the cloud holds a neural network and the client holds an input to the network, typically an image and the communication happens in the encrypted domain.
Homomorphic encryption, which can be used to implement SMC, is also used for privacy protection in neural networks.
This cryptographic technique allows (all or a subset of) operations to be performed on the encrypted data without the need for decryption. 
These works~\cite{liu17ccs,dowlin16icml} suggest the client/edge device encrypts the data (on top of the communication encryption, e.g., SSL) before sending it to the cloud; which it then, performs operations on the encrypted data and returns the output. 
Nevertheless, this approach suffers from a prohibitive computational and communication cost, exacerbating the complexity and compute-intensivity of neural networks especially on resource-constrained edge devices.
\shredder, in contrast, avoids the significant cost of encryption or homomorphic data processing.

\begin{table}
    
    \caption{Privacy Protecting methods in DNNs.}
    \label{tab:related}
    
    {\centering
\begin{tabular}{@{}l|l|l@{}}
\toprule
\label{tab:related}
	& \small Non-Intrusive & \small  Intrusive \\
\midrule
\multirow{2}{*}{Inference} & \textit{\textbf{\sffamily\small  Shredder}} & \small CryptoNets~\cite{dowlin16icml}, GAZELLE~\cite{juvekar18usenix} \\
& \small MiniONN~\cite{liu17ccs} & \small Arden~\cite{wang18kdd}, DPFE~\cite{osia2017hybrid,osia18kde}\\
\midrule
\multirow{2}{*}{ \small Training/DB} &  \small Rappor~\cite{erlingsson14ccs} &  \small With Differential Privacy~\cite{shokri15ccs,abadi16DeepLW},\\
&  \small Apple~\cite{apple17white} &  \small SecureML~\cite{mohassel17sp}\\
\bottomrule
\end{tabular}
}
\vspace{-1ex}    
\end{table}

\vspace{-1ex}
\section{Conclusion}
Privacy is a fundamental human right recognized in the United Nations (UN) Declaration of Human Rights.
However, the systems and computational infrastructure in use seem to have been designed for offering functionality without foundational consideration for privacy. 
Such a gap is more concerning today since cloud-based deep learning service make their way to the households as home automation devices or shape our social, political, and economical interactions.
A single paper is not an answer to this brewing epidemic, but it represents an initial effort to build a mathematically-sound private systems that offer reliable degrees of utility.
As such, this paper examines the use of noise to reduce the information content of the communicated data to the cloud while still maintaining high levels of accuracy.
By casting the noise injection as a learning process that uses differentiation to find the distribution of the noise, we devise \deepmist, which strikes an asymmetric balance between accuracy and privacy with formal mathematical guarantees.
Experimentation with multiple real-life DNNs showed that \shredder can significantly reduce the information content of the communicated data with only \avgaccloss accuracy loss.
These results pave a promising path forward.
\section{Acknowledgment}

We thank the anonymous reviewers for their insightful comments.
This work was in part supported by National Science Foundation (NSF) awards CNS\#1703812, ECCS\#1609823, CCF\#1553192, 
Air Force Office of Scientific Research (AFOSR) Young Investigator Program (YIP) award \#FA9550-17-1-0274,  
National Institute of Health (NIH) award \#R01EB028350, and
%
Air Force Research Laboratory (AFRL) and Defense Advanced Research Project Agency (DARPA) under agreement number \#FA8650-20-2-7009 and \#HR0011-18-C-0020.
The U.S. Government is authorized to reproduce and distribute reprints for Governmental purposes notwithstanding any copyright notation thereon.
The views and conclusions contained herein are those of the authors and should not be interpreted as necessarily representing the official policies or endorsements, either expressed or implied, of Arm, Amazon, NSF, AFSOR, NIH, AFRL, DARPA or the U.S. Government.

\bibliographystyle{ieeetr}
\bibliography{ref}

\end{document}